\documentclass[10pt,a4paper,onecolumn,twoside]{article}
\usepackage[top=2cm,bottom=2cm,left=2cm,right=2cm]{geometry}

\usepackage[utf8]{inputenc}
\usepackage[T1]{fontenc}

\usepackage{fourier}

\usepackage{amsmath}
\allowdisplaybreaks
\usepackage{booktabs}
\usepackage{graphicx}

\usepackage{amsfonts}
\usepackage{amssymb}
\usepackage{stmaryrd}
\usepackage{mathrsfs}
\usepackage{bm}

\usepackage{amsthm}
\theoremstyle{remark}

\theoremstyle{theorem}
\newtheorem*{condition}{Condition}

\usepackage{framed}
\usepackage[hidelinks]{hyperref}

\title{Nonlinear plane waves in saturated porous media\\ with incompressible constituents}
\author{Harold Berjamin\\ \\
	\emph{\footnotesize School of Mathematics, Statistics and Applied Mathematics,}\\ \emph{\footnotesize NUI Galway, University Road, Galway, Republic of Ireland}}
\date{}

\usepackage[numbers,sort&compress]{natbib}
\begin{document}

\pagestyle{headings}
\thispagestyle{empty}
\maketitle

\begin{abstract}
	\noindent
	We consider the propagation of nonlinear plane waves in porous media within the framework of the Biot--Coussy biphasic mixture theory. The tortuosity effect is included in the model, and both constituents are assumed incompressible (Yeoh-type elastic skeleton, and saturating fluid). In this case, the linear dispersive waves governed by Biot's theory are either of compression or shear-wave type, and nonlinear waves can be classified in a similar way. In the special case of a neo-Hookean skeleton, we derive the explicit expressions for the characteristic wave speeds, leading to the hyperbolicity condition. The sound speeds for a Yeoh skeleton are estimated using a perturbation approach.
	Then we arrive at the evolution equation for the amplitude of acceleration waves.
	In general, it is governed by a Bernoulli equation. With the present constitutive assumptions, we find that longitudinal jump amplitudes follow a nonlinear evolution, while transverse jump amplitudes evolve in an almost linearly degenerate fashion. \\ 
	
\end{abstract}

\section{Introduction}\label{sec:Intro}

Originating in the field of geophysics, the theory of porous media has a long history that goes back to the 18th century (see the historical review by De Boer \cite{deboer96a}).
Poroelasticity theories have also been employed in various biomechanical applications involving the deformation of hydrated porous biological tissues. As noted by Ateshian \cite{ateshian17}, one early application to biological tissues is the modelling of articular cartilage. More recently, multiphasic models have been used to model the mechanical response of brain tissue, which is known to be very soft, heterogeneous, nonlinear, and time-dependent (see the review by Budday et al. \cite{budday19}). Based on quasi-static mechanical loadings, recent laboratory studies show that the fluid-solid coupling in brain tissue
may be partly responsible for time-dependent effects \cite{forte17, hosseinifarid20, comellas20, wang19}. Consequently, biphasic theory is receiving increasing attention in brain mechanics, where it has also been used for the modelling of drug delivery and surgical procedures \cite{ehlers14, forte18}.

Up to now, biphasic brain material models have mainly been used in quasi-static configurations. Efforts to address dynamic problems are strongly motivated by the study of traumatic events, in particular Traumatic Brain Injury (TBI) \cite{suer2020}.
To increase our understanding of head trauma, a major challenge lies in the development and validation of computational models, for which the determination of parameters is a key prerequisite \cite{jiang15}.
Ranging from mild injuries to severe concussions, head traumas involve a large range of wave amplitudes and frequencies.
Thus, neither linear material models (valid at small amplitudes) nor purely elastic models (valid at low frequencies) are satisfactory.
Instead, all-solid nonlinear viscoelasticity models have been successfully used to simulate traumatic events \cite{trotta20, tripathi19}.
Nevertheless, in some cases the cerebrospinal fluid (CSF) which hydrates the brain has been shown to play a crucial mechanical  role, more precisely due to cavitation phenomena occurring at high pressure levels \cite{panzer12}.
Since time-dependent effects such as fluid-solid couplings are decisive in dynamic configurations, it is reasonable to consider that poroelasticity might play an important role in TBI as well.

Motivated by the above-mentioned observations, the present study aims at gaining insight in the wave physics of nonlinear porous materials by estimating wave speeds and amplitudes analytically. Such results are quite rare in the nonlinear mixture theory literature, where most analytical results have been obtained in the linear limit \cite{deboer94, deboer95, deboer96b}. In particular, the proposed fully nonlinear analysis is of interest for the validation of computational methods \cite{diebels96,deboer96b,breuer99,heider12}. A strongly related study is that by Ciarletta et al. \cite{ciarletta18}, where the decay of nonlinear acceleration waves is investigated. Arising in a very different context, less related works encompass also the computation of nonlinear compaction waves in geophysical porous media \cite{connolly15}.

As far as mechanical modelling is concerned, mainly two complementary approaches are found in the literature. A first approach consists in extending the linear Biot theory \cite{bourbie87, carcione15} to finite strain, by using the same quasi-variational formalism along with a hyperelastic strain energy function. In particular, this `ad hoc' approach has been used in relation with geophysical wave motion \cite{grinfeld96,tong17}. The same strategy was also followed by Ciarletta et al. \cite{ciarletta18} in their study of nonlinear acceleration waves. A second approach known as \emph{mixture theory} derives from the ground principles of continuum mechanics, namely balance principles and thermodynamic restrictions. This `rational' approach has been used in various biomechanical applications \cite{ateshian17}, among which some of the most recent quasistatic brain mechanics studies \cite{forte17, comellas20, hosseinifarid20}.

While the traditional mixture modelling approach is appealing, these theories are not consistent with the linear Biot theory in the infinitesimal strain limit \cite{deboer05}.
Known as the Biot--Coussy theory, Coussy's modified mixture theory \cite{coussy04} enables direct links with the linear Biot theory.
Based on these works, here we introduce a dynamic biphasic model with incompressible constituents that includes Biot's tortuosity effect. The skeleton is assumed non-linear elastic with Yeoh-type behaviour, and viscous effects are neglected in the fluid's partial stress. Various connections with existing models and anterior works are identified.
However, it is not known yet if the present theory is valid to model head trauma in its present form, due to the current lack of experimental data in TBI-related settings. Here, the parameter values had to be inferred from quasi-static experiments \cite{comellas20}.

Despite this apparent practical limitation, the study uncovers several general analytical results about the propagation of nonlinear waves in porous media.
Starting with travelling plane waves, the characteristic wave speeds and the decay of acceleration waves are then investigated. The conditions of hyperbolicity follow from the requirement of real wave speeds. Under the present constitutive assumptions, longitudinal jump amplitudes are shown to satisfy a nonlinear evolution, while transverse jump amplitudes decay in a quasi linearly degenerate fashion.

The paper is organised as follows. In Section~\ref{sec:Theory}, the main equations of biphasic mixture theory are presented. In Section~\ref{sec:Biot}, the governing equations are linearised, and the dispersion properties of infinitesimal Biot waves are recalled. The main results are presented in Section~\ref{sec:nlin}, where nonlinear plane wave motion is considered. Prospective future works are discussed in the conclusion part (Section \ref{sec:Conclu}).

%%%%%%%%%%%%% Mixture %%%%%%%%%%%%%%%

\section{Biphasic mixture theory}\label{sec:Theory}

%%%%%%%%%%%%%%%%%%%%%%%%%%%%%%%%

We consider an unbounded fluid-saturated porous material, to be described within the framework of the theory of porous media (Biot--Coussy biphasic mixture theory). In what follows, we introduce the main equations describing the motion and the deformation of such a fluid-solid mixture. The solid skeleton is assumed elastic, and heat transfer is neglected. Here, several shortcuts are taken for the sake of conciseness. For more details, the reader is referred to various reference textbooks \cite{bowen76,rajagopal95,ehlers02,coussy04,deboer05} and other related works \cite{macminn16,ateshian17}.

%+++++++++++++++++++++++

\subsection{Kinematics}

%+++++++++++++++++++++++

Consider the position $\bm{x}$ of a particle. Its components are expressed with respect to an orthonormal basis $(\bm{e}_1,\bm{e}_2,\bm{e}_3)$ of the Euclidean space, and a Cartesian coordinate system is chosen. The Eulerian position vector $\bm x$ is the same for fluid particles and solid particles, but the reference position vectors $\bm{X}^\text{f}$, $\bm{X}^\text{s}$ for the fluid and solid phases are independent. Let $n^\alpha$ denote the Eulerian volume fraction of the phase $\alpha \in \lbrace\text{s}, \text{f}\rbrace${\,---\,}i.e., the volume fraction of the solid and fluid constituent in the deformed configuration. The saturation condition requires
\begin{equation}
	n^\text{s} + n^\text{f} = 1 \, .
	\label{Saturation}
\end{equation}
The fluid volume fraction $n^\text{f}$ corresponds to the Eulerian porosity.

In the Eulerian description  of motion, spatial differential operators computed with respect to $\bm{x}$ are written $\mathrm{div}$, $\mathrm{grad}$, etc. In the skeleton-Lagrangian description of motion, the spatial coordinate is the position $\bm{X}^\text{s}$ of a skeleton particle in its reference (undeformed) configuration. Spatial differential operators computed with respect to $\bm{X}^\text{s}$ are written $\mathrm{Div}$, $\mathrm{Grad}$, etc. 

We introduce the deformation gradient tensor $\bm{F} = \text{Grad}\,{\bm{x}}$ of the solid phase, as well as its inverse $\bm{A} = \bm{F}^{-1}$ given by $\bm{A} = \text{grad}\, \bm{X}^\text{s}$. By introducing the displacement field $\bm{u}^\text{s} = \bm{x}-\bm{X}^\text{s}$ of the solid phase, the deformation gradient tensor and its inverse are rewritten
\begin{equation}
	\bm{F} = \bm{I} + \mathrm{Grad}\, \bm{u}^\text{s} ,\qquad
	\bm{A} = \bm{I} - \mathrm{grad}\, \bm{u}^\text{s} ,
	\label{F}
\end{equation}
where $\bm{I} = [\delta_{ij}]$ is the metric tensor, here represented by Kronecker delta components $\delta_{ij}$. In related works, the tensor $\bm{A}$ is called the \emph{distorsion tensor} \cite{romenski20,godunov03}. Various strain tensors can be defined as functions of $\bm F$ or $\bm{A}$, such as the right Cauchy–Green deformation tensor $\bm{C} = \bm{F}^{\sf T}\! \bm{F}$ and the Green--Lagrange strain  tensor $\bm{E} = \frac12 (\bm{C}-\bm{I})$.

To describe the mixture's motion, we introduce the velocity fields $\bm{v}^\alpha = \bm{x}'_\alpha$. The prime with index $\alpha$ denotes the particle time derivative, which is computed while following the motion of the solid ($\alpha=\text{s}$) or of the fluid ($\alpha=\text{f}$). Thus, for any scalar Eulerian field $\Gamma(\bm{x},t)$, we have
\begin{equation}
	\Gamma'_\alpha = \frac{\partial \Gamma}{\partial t} + (\text{grad}\, \Gamma) \bm{\cdot v}^\alpha ,
	\qquad
	\alpha \in \lbrace\text{s}, \text{f}\rbrace ,
	\label{MaterialDiff}
\end{equation}
and similar differentiation operators can be introduced for vectorial and tensorial fields. The velocity fields may be rewritten as $\bm{v}^\alpha = (\bm{u}^\alpha)'_\alpha$ where $\bm{u}^\alpha = \bm{x} - \bm{X}^\alpha$ denotes the displacement from a particle of $\alpha$ from its reference position $\bm{X}^\alpha$ to its current position $\bm x$.

We also introduce  the Eulerian velocity gradients $\bm{L}^\alpha = \text{grad}\, \bm{v}^\alpha$ and their symmetric part $\bm{D}^\alpha = \frac12 (\bm{L}^\alpha + \bm{L}^{\alpha{\sf T}})$.
In the solid phase, the velocity gradient $\bm{L}^\text{s} = \bm{F}'_\text{s} \bm{F}^{-1}$ depends on the deformation gradient and its material derivative, and so does the symmetric part $\bm{D}^\text{s} = \bm{F}^{-{\sf T}}\! \bm{E}'_\text{s} \bm{F}^{-1}$.
Finally, to describe fluid motion with respect to the skeleton, we introduce the seepage velocity $\bm{w} = \bm{v}^\text{f} - \bm{v}^\text{s}$ such that $\Gamma'_\text{f} = \Gamma'_\text{s} + (\text{grad}\, \Gamma) \bm{\cdot w}$.

%+++++++++++++++++++++++

\subsection{Balance principles}

%++++++++++++++++++++++

Let us neglect mass transfer and external mass supply. We introduce the mass densities $\rho^\alpha = n^\alpha\!\rho^{\alpha\text{R}}$, where $\rho^{\alpha\text{R}}$ are the real mass densities. Also known as true, intrinsic or effective material density, $\rho^{\alpha\text{R}}$ represents the mass of a constituent per volume of that constituent. 

In this study, we consider an incompressible skeleton saturated by an incompressible fluid. By definition, the true density $\rho^{\alpha\text{R}}$ of an incompressible constituent is invariant. Under this assumption, the Eulerian mass continuity equation for each constituent $\alpha \in \lbrace\text{s}, \text{f}\rbrace$ reads
\begin{equation}
	(n^\alpha)'_\alpha + n^\alpha \mathrm{div}\, \bm{v}^\alpha = 0 \, ,
	\label{ContinuityEIncomp}
\end{equation}
where the material derivative is defined in Eq.~\eqref{MaterialDiff}. By summation of both continuity equations, the saturation constraint \eqref{Saturation} yields the condition
\begin{equation}
	\text{div} \big(n^\text{f}\bm{v}^\text{f} + n^\text{s}\bm{v}^\text{s}\big) = 0 \, ,
	\label{ConstIncomp1}
\end{equation}
which will be used to enforce saturation later on.

Introducing the volume dilatation $J = \det\bm{F}$ of the solid phase, the relation
$J = {n_0^\text{s}}/{n^\text{s}}$ is obtained by integration of the solid mass balance equation of Eq.~\eqref{ContinuityEIncomp}, where $n^\text{s}_0 = 1-n^\text{f}_0$ denotes the volume fraction of the solid phase in the reference configuration. Therefore, porosity $n^\text{f} = 1 - n^\text{s}_0/J$ is function of the deformation. Contrary to monophasic incompressible solids which support only isochoric deformations (i.e., $J\equiv 1$ is prescribed), the volume dilatation of biphasic particles increases with porosity, in a similar fashion to a squeezed sponge.

The balance of linear momentum equation for each constituent reads
\begin{equation}
	\rho^\alpha (\bm{v}^\alpha)'_\alpha = \mathrm{div}\, \bm{\sigma}^\alpha + \rho^\alpha \bm{b}^\alpha + \hat{\bm p}^\alpha ,
	\label{ConsMomMult}
\end{equation}
where the vector $\bm{b}^\alpha$ denotes external body forces per unit mass.
The reciprocity condition $\hat{\bm p}^\text{s} = -\hat{\bm p}^\text{f}$ follows from the balance of linear momentum applied to the mixture as a whole.
Assuming microscopically non-polar constituents, the symmetry of the partial Cauchy stresses $\bm{\sigma}^\alpha = \bm{\sigma}^{\alpha{\sf T}}\!$ is deduced from the balance of moment of momentum, where no supply of momentum is included.

In the absence of heat flux (adiabatic case), energy transfer and external energy sources, the local form of the balance of internal energy for each constituent reads
\begin{equation}
	\rho^\alpha (e^\alpha)'_\alpha = \bm{\sigma}^\alpha:\bm{D}^\alpha - \hat{\bm p}^\alpha\!  \bm{\cdot v}^\alpha ,
	\label{EnergyEMult}
\end{equation}
where $e^\alpha$ denotes the specific internal energy. The colon denotes the double contraction of second-order tensors. Introducing the density of mechanical energy $\mathcal{E}^\alpha = e^\alpha + \frac12 \|\bm{v}^\alpha\|^2$, one may express the balance of mechanical energy as
\begin{equation}
	\rho^\alpha (\mathcal{E}^\alpha)'_\alpha = \mathrm{div}(\bm{\sigma}^\alpha \bm{v}^\alpha) + \rho^\alpha\bm{b}^\alpha\!  \bm{\cdot v}^\alpha,
	\label{EnergyMechMult}
\end{equation}
for each constituent. Here, $\bm{\sigma}^\alpha \bm{v}^\alpha$ represents the Poynting vector, and $\rho^\alpha \bm{v}^\alpha\!  \bm{\cdot b}^\alpha$ is the work done by the external body force.

For consistency with Biot's linear theory of saturated porous media, some adjustments have to be made \cite{coussy04}. Indeed, the local balance of energy \eqref{EnergyEMult} over the fluid phase $\alpha = \text{f}$ does not include the `tortuosity' effect of Biot's theory, which cannot be captured by the macroscopic mixture approach. Following estimations at the scale of a representative volume, Coussy \cite{coussy04} introduces the \emph{tortuosity vector} $\bm{ a} = (a-1)\, \bm{w}'_\text{f}$ that modifies the balance of energy \eqref{EnergyEMult}-\eqref{EnergyMechMult} for the fluid phase as follows:
\begin{equation}
	\begin{aligned}
		\rho^\text{f} (e^\text{f})'_\text{f} &= \bm{\sigma}^\text{f}:\bm{D}^\text{f} - \hat{\bm p}^\text{f}\bm{\cdot v}^\text{f} - \rho^\text{f}\bm{a}  \bm{\cdot w} \, ,\\
		\rho^\text{f} (\mathcal{E}^\text{f})'_\text{f} &= \mathrm{div}(\bm{\sigma}^\text{f} \bm{v}^\text{f}) + \rho^\text{f}\bm{b}^\text{f}  \bm{\cdot v}^\text{f} - \rho^\text{f}\bm{a}  \bm{\cdot w} \, .
	\end{aligned}
	\label{EnergyTort}
\end{equation}
The tortuosity factor $a \geq 1$ is defined as the ratio between the seepage energy averaged over an elementary volume, and the corresponding macroscopic quantity $\rho^\text{f} \|\bm{w}\|^2$. It satisfies $a \to 1$ in the single-constituent fluid limit, and $a\to +\infty$ in the single-constituent solid limit. As noted by Wilmanski \cite{wilmanski05}, objective relative accelerations may be introduced instead of the relative acceleration $\bm{w}'_\text{f}$ to account for the tortuosity effect. We will see later on that the tortuosity vector of Eq.~\eqref{EnergyTort} adds $-\rho^\text{f}\bm{a}$ to the interaction force $\hat{\bm p}^\text{f}$, leading to a Cattaneo-type effect on the filtration law. The simple mixture model without tortuosity effect is recovered by setting $a\equiv 1$.

\subsection{Constitutive modelling}

In contrast to the above balance principles which are written for each constituent, the postulate of entropy increase is written for the mixture as a whole. 
In a standard way, we consider a single-temperature mixture  such that $\theta>0$ is the temperature field for all the constituents, and $\eta^\alpha$ are the specific entropies. Thus, the second principle of thermodynamics is expressed by the Clausius--Duhem inequality
\begin{equation}
	\mathscr{D} = \rho^\text{s}\theta (\eta^\text{s})'_\text{s} + \rho^\text{f}\theta (\eta^\text{f})'_\text{f}  \geq 0 \, ,
	\label{CD1}
\end{equation}
where $\mathscr{D}$ is the dissipation in the mixture.

The thermodynamic procedure based on the temperature is well-described in the literature \cite{ehlers02,deboer05}. To model nearly isentropic processes such as acoustic perturbations \cite{pierce19}, one may assume that the biphasic mixture is described by the state variables $\lbrace \eta^\text{s} , \eta^\text{f}, \bm{E}\rbrace$, where $\bm E$ is the Green--Lagrange strain tensor. Moreover, because we are considering a constrained mixture with incompressible constituents, various simplifications can be performed \cite{ehlers02}. Here, phase separation is assumed, which amounts to stipulate that $e^\text{s}$ is function of $\lbrace\eta^\text{s}, \bm{E}\rbrace$, and that $e^\text{f}$ is function of $\eta^\text{f}$ only. Thus, according to the Gibbs identity, the total material derivatives of the functions of state $e^\alpha$ satisfy
\begin{equation}
	\begin{aligned}
		&\rho^\text{s} (e^\text{s})'_\text{s}
		- \rho^\text{s} \frac{\partial e^\text{s}}{\partial \eta^\text{s}} (\eta^\text{s})'_\text{s} - \rho^\text{s} \frac{\partial e^\text{s}}{\partial \bm{E}} : \bm{E}'_\text{s} = 0 \, ,
		\\
		&\rho^\text{f} (e^\text{f})'_\text{f}
		- \rho^\text{f} \frac{\partial e^\text{f}}{\partial \eta^\text{f}} (\eta^\text{f})'_\text{f} = 0
		\, .
	\end{aligned}
	\label{Gibbs}
\end{equation}
The intrinsic incompressibility of the constituents is introduced using the method of Lagrange multipliers. For this purpose, the differential form \eqref{ConstIncomp1} of the saturation constraint is expanded as follows using vector calculus identities:
\begin{equation}
	p \left( n^\text{s}\text{div}\, \bm{v}^\text{s} + n^\text{f}\text{div}\, \bm{v}^\text{f} + (\text{grad}\, n^\text{f}) \bm{\cdot w} \right) = 0 \, ,
	\label{ConstIncomp2}
\end{equation}
where the corresponding Lagrange multiplier $p$ has been introduced.

Using the conservation of energy \eqref{EnergyEMult}, summation of the above equations \eqref{CD1}-\eqref{ConstIncomp2} yields the final expression of the dissipation.
Due to the tortuosity effect of Eq.~\eqref{EnergyTort}, the dissipation becomes
\begin{equation}
	\begin{aligned}
		\mathscr{D} = \left(\bm{\sigma}^\text{s} + n^\text{s} p\bm{I} - \frac{1}{J}\bm{F}\frac{\partial W}{\partial \bm{E}}\bm{F}^{\sf T}\right):\bm{D}^\text{s} %\\ &\phantom{ = }
		+ (\bm{\sigma}^\text{f} + n^\text{f}p\bm{I}):\bm{D}^\text{f} - (\hat{\bm p}^\text{f} + \rho^\text{f}\bm{a} - p\, \text{grad}\, n^\text{f}) \bm{\cdot w} \, ,
		\label{CD2}
	\end{aligned}
\end{equation}
where $\theta = {\partial e^\alpha}/{\partial \eta^\alpha}$ is required to ensure the positivity of the dissipation for arbitrary transformations. Here, we have used the reciprocity condition $\hat{\bm p}^\text{s} = -\hat{\bm p}^\text{f}$, and we have introduced the strain energy density function $W = \rho^\text{s}_0 e^\text{s}$ of the skeleton, with $\rho^\text{s}_0 = n^\text{s}_0 \rho^\text{sR}$.
Following standard arguments, the dissipation inequality \eqref{CD1} is satisfied for arbitrary processes if
\begin{equation}
	\begin{aligned}
		\bm{\sigma}^\text{s} &= -n^\text{s}p\bm{I} + \frac{1}{J}\bm{F}\frac{\partial W}{\partial \bm{E}}\bm{F}^{\sf T} , \quad&
		\bm{\sigma}^\text{f} &= -n^\text{f}p\bm{I} \, , \\
		\hat{\bm p}^\text{f} &= -\rho^\text{f}\bm{a} + p\, \text{grad}\, n^\text{f} + \hat{\bm p}^\text{f}_\text{e} \, , \quad&
		\mathscr{D} &= -\hat{\bm p}^\text{f}_\text{e}  \bm{\cdot w} \geq 0 \, .
	\end{aligned}
	\label{Constitutive}
\end{equation}
Known as the effective drag force, the quantity $\hat{\bm p}^\text{f}_\text{e}$ entails no dissipation if orthogonal to $\bm w$.

Using the saturation condition \eqref{Saturation}, Terzaghi's effective Cauchy stress reads
\begin{equation}
	\bm{\sigma}^\text{e} = {\bm \sigma}^\text{i} + p \bm{I} = \frac{1}{J} \bm{F} \frac{\partial W}{\partial \bm{E}} \bm{F}^{\sf T} ,
	\label{Terzaghi}
\end{equation}
where ${\bm \sigma}^\text{i} = {\bm \sigma}^\text{s} + {\bm \sigma}^\text{f}$ denotes the inner part of the mixture stress. As explained by Carcione \cite{carcione15}, ``the effective-stress concept means that the response of the saturated porous medium is described by the response of the dry porous medium with the applied stress replaced by the effective stress''.

The remaining dissipation $\mathscr{D}$ in Eq.~\eqref{Constitutive} is ensured positive by setting
\begin{equation}
	\hat{\bm p}^\text{f}_\text{e} = -\frac{(n^\text{f})^2}{k^\text{f}} \bm{w} \, ,
	\label{Drag}
\end{equation}
which models the internal friction between solid and fluid. The parameter $k^\text{f} \geq 0$ is the permeability of the fluid, i.e. the ratio of the skeleton's intrinsic permeability and the fluid's dynamic viscosity. It satisfies $k^\text{f} \to +\infty$ in the single-constituent fluid limit, and $k^\text{f}\to 0$ in the single-constituent solid limit. Injecting the expression of $\hat{\bm p}^\text{f}_\text{e}$ in the conservation of momentum equation \eqref{ConsMomMult} for the fluid constituent, one eventually obtains \emph{Darcy's filtration law},
\begin{equation}
	n^\text{f}\bm{w} = -k^\text{f} \left[\text{grad}\, p - \rho^\text{fR}\big(\bm{b}^\text{f} - (\bm{v}^\text{f})'_\text{f} - \bm{a}\big)\right] .
	\label{Darcy}
\end{equation}
More general forms of Darcy's law may include a permeability tensor instead of the scalar $k^\text{f}$.

\paragraph{Remark.}
%\begin{remark}
	In the case of fluid flow through a rigid porous skeleton, the porosity $n^\text{f}$ is constant. The interaction force of Eqs.~\eqref{Constitutive}-\eqref{Drag} becomes
	\begin{equation}
		\hat{\bm p}^\text{f} = -\rho^\text{f}\bm{a} + \hat{\bm p}^\text{f}_\text{e} = -\frac{(n^\text{f})^2}{k^\text{f}} \left( \tau_a \bm{w}'_\text{f} + \bm{w} \right) ,
		\label{Cattaneo}
	\end{equation}
	where we have used the definition $\bm{ a} = (a-1)\, \bm{w}'_\text{f}$, and $\tau_a = (a-1) k^\text{f}\rho^\text{fR}/n^\text{f}$ is a characteristic time. Thus, we note that the tortuosity effect of Eq.~\eqref{EnergyTort} yields a Cattaneo-type relaxation in the filtration law. Again, one may have replaced the present relative acceleration $\bm{w}'_\text{f}$ by an objective derivative \cite{wilmanski05}, e.g. in a similar fashion to the so-called Darcy--Jordan--Cattaneo model of Ref.~\cite{jordan17}. \smallskip
%\end{remark}

We assume that the skeleton's effective mechanical response \eqref{Terzaghi} follows from the two-term Yeoh strain energy function
\begin{equation}
	W = \tfrac12 \mu \left((I_1 - 3) + \tfrac12\beta (I_1 - 3)^2 - 2 \ln J\right) + \tfrac12\lambda (\ln J)^2 ,	\label{NeoHookean}
\end{equation}
where $I_1 = \text{tr}\, \bm{B}$ is the first principal invariant of the left Cauchy--Green tensor $\bm{B} = \bm{F}\bm{F}^{\sf T}\!$. The corresponding constitutive relation reads
\begin{equation}
	J\bm{\sigma}^\text{e} = \mu \big( \bm{B} + \beta (I_1-3) \bm{B} - \bm{I}\big) + \lambda (\ln J) \bm{I} ,
\end{equation}
where the positive constants $\lambda$, $\mu$ are the Lam{\'e} parameters of linear elasticity. The Yeoh parameter $\beta \geq 0$ has been introduced for sake of generality, in view of discussing the influence of the constitutive assumptions. With this choice, the neo-Hookean model $\beta \to 0$ used by Diebels and Ehlers \cite{diebels96} is recovered as a special case. While the analysis introduced hereinafter is quite general, most of the exact analytical formulas are obtained in the neo-Hookean limit. The more general case $\beta > 0$ is addressed in a quasi-analytical fashion.

Moreover, we assume that the fluid's permeability $k^\text{f}$ follows from the formula \cite{markert07}
\begin{equation}
	k^\text{f} = k^\text{f}_0 \left(\frac{n^\text{f}}{n^\text{f}_0} \frac{1-n^\text{f}_0}{1-n^\text{f}}\right)^{\kappa} ,
	\label{Permeability}
\end{equation}
which is an alternative to the Kozeny--Carman formula of Refs.~\cite{macminn16, coussy04}. Here, $k^\text{f}_0$ represents the fluid's permeability when the porosity $n^\text{f}$ equals its initial value $n^\text{f}_0$, and $\kappa$ is a dimensionless parameter. Lastly, we assume that the tortuosity coefficient $a$ satisfies Berryman's formula \cite{coussy04, wilmanski05, carcione15}
\begin{equation}
	a = \frac12 \left( 1 + \frac{1}{n^\text{f}}\right) ,
	\label{Tortuosity}
\end{equation}
but more general expressions could be used \cite{markert07}.

For the purpose of illustration, typical values of the material parameters for a soft biological tissue saturated by an incompressible liquid are specified in Table~\ref{tab:Params}. The elastic parameters $\lambda$, $\mu$ are deduced from Comellas et al. \cite{comellas20},\footnote{The numerical value of the Ogden parameter $\mu_2$ in Ref.~\cite{comellas20} has a typo, see Franceschini et al. \cite{franceschini06}.} and the parameter $\beta$ has been chosen in such a way that shear stresses are consistent with Ref.~\cite{comellas20} over a large range of deformations (simple shear strains ranging from $-0.7$ to $0.7$). The tortuosity coefficient \eqref{Tortuosity} deduced from the reference value of the porosity $n^\text{f}_0$ is $a = 3.0$.

\begin{table}
	\centering
	\caption{Physical parameters of water-saturated brain tissue inferred from Ref.~\cite{comellas20}, where the mass density of water at room temperature is assumed for both constituents $\alpha \in \lbrace\text{s},\text{f}\rbrace$. The potential mismatch between isothermal and isentropic measurements is neglected in the present study. \label{tab:Params}}
	
	\vspace{0.2em}
	{\renewcommand{\arraystretch}{1.1}
		\setlength{\tabcolsep}{3pt}
		\begin{tabular}{ccccccc}
			\toprule
			$\lambda$ [kPa] & $\mu$ [kPa] & $\beta$ &  $\rho^{\alpha\text{R}}$ [kg/m\textsuperscript{3}] & $n_0^\text{f}$ & $k_0^\text{f}$ [m\textsuperscript{2}/(Pa.s)] & $\kappa$ \\
			$334$ & $6.82$ & $2.2$ & $997$ & $0.20$ & $8.9 \times 10^{-14}$ & $40$ \\
			\bottomrule
	\end{tabular}}
\end{table}

\subsection{Eulerian equations of motion}

In the Eulerian specification of motion,\footnote{The skeleton-Lagrangian form of the equations of motion is described in Li et al. \cite{li04}, see also Wilmanski \cite{wilmanski05}.} spatial differential operators are computed with respect to $\bm x$. We consider a fluid-saturated poroelastic material with incompressible constituents governed by Eqs.~\eqref{ContinuityEIncomp}-\eqref{ConsMomMult}. In addition, a kinematic relationship between the distorsion tensor $\bm{A}$ defined in Eq.~\eqref{F} and the velocity $\bm{v}^\text{s}$ of the solid phase is introduced in the first line of Eq.~\eqref{HypSmall} below. The latter can be retrieved by using the equality of mixed partials in Eq.~(8.3) of Godunov and Romenskii \cite{godunov03}.
Thus, the equations of motion read as a system of balance laws constrained by the saturation condition of Eq.~\eqref{Saturation}.
This system is closed by the constitutive equations for the partial stresses $\bm{\sigma}^\alpha$ and the interaction forces $\hat{\bm p}^\text{f} = -\hat{\bm p}^\text{s}$ in Eqs.~\eqref{Constitutive}-\eqref{Drag}. We therefore end up with a system of eighteen scalar equations, which involves the eighteen components of $\lbrace \bm{A} , n^\alpha, \bm{v}^\alpha, p \rbrace$ for $\alpha \in \lbrace\text{s}, \text{f}\rbrace$.

Keeping Eqs.~\eqref{ContinuityEIncomp}-\eqref{ConsMomMult} for the fluid phase and adding the latter to the equations for the solid phase, we may rewrite the above system as
\begin{gather}
	\left\lbrace
	\begin{aligned}
		&\partial_t \bm{A} + \text{grad} (\bm{A} \bm{v}^\text{s}) = \bm{0}, \\
		&\partial_t n^\text{f} + \text{div} \big(n^\text{f}(\bm{w} + \bm{v}^\text{s}) \big) = 0, \\
		&\text{div} \big(n^\text{f}\bm{w} + \bm{v}^\text{s} \big) = 0, \\
		&\rho^\text{f}\left[ \partial_t \bm{v}^\text{s} + (\text{grad}\, \bm{v}^\text{s})(\bm{w} + \bm{v}^\text{s}) \right] %\\ &
		+ a\rho^\text{f}\left[\partial_t \bm{w} + (\text{grad}\, \bm{w}) (\bm{w} + \bm{v}^\text{s})\right] % \\ &\qquad 
		= -n^\text{f}\mathrm{grad}\, p - \tfrac{(n^\text{f})^2}{k^\text{f}} \bm{w} + \rho^\text{f} \bm{b}^\text{f}, \\
		&\rho^\text{s} \left[ \partial_t \bm{v}^\text{s} + (\text{grad}\, \bm{v}^\text{s})\bm{v}^\text{s} \right] + \rho^\text{f} \left[\partial_t (\bm{w} + \bm{v}^\text{s}) + \text{grad}(\bm{w} + \bm{v}^\text{s})(\bm{w} + \bm{v}^\text{s})\right] %\\&\qquad 
		= \mathrm{div}\, \bm{\sigma}^\text{i} + \rho\bm{b},
	\end{aligned}
	\right.
	\label{HypSmall} \raisetag{4\baselineskip}
\end{gather}
which introduces the mixture's inner stress $\bm{\sigma}^\text{i} = \bm{\sigma}^\text{e} - p\bm{I}$, see Eq.~\eqref{Terzaghi}, effective body force $\rho\bm{b} = \rho^\text{s} \bm{b}^\text{s} + \rho^\text{f} \bm{b}^\text{f}$, and effective density $\rho = \rho^\text{s} + \rho^\text{f}$. We thus end up with a system of seventeen scalar equations, which involves the seventeen components of $\lbrace \bm{A} , n^\text{f}, \bm{v}^\text{s}, \bm{w}, p \rbrace$, where $\bm{w} = \bm{v}^\text{f} - \bm{v}^\text{s}$ is the seepage velocity. For sake of exhaustiveness, let us mention that appropriate boundary conditions should be provided \cite{coussy04}. Here, plane waves propagating in unbounded domain are considered.

Note that standard vector calculus identities can be used to derive alternative forms. In particular, the second line of Eq.~\eqref{HypSmall} might be removed, since the porosity $n^\text{f}$ is function of $\bm A$ (consequence of Eq.~\eqref{ContinuityEIncomp}). Moreover, the last line of Eq.~\eqref{HypSmall} may be rewritten in more compact form by introducing the mixture velocity and the mixture stress tensor, see Refs.~\cite{rajagopal95, ateshian17}. Contrary to the case $a \equiv 1$ of simple mixtures, no fully conservative first-order formulation of the equations of motion is known due to the dependency of the tortuosity coefficient \eqref{Tortuosity} with porosity.
When $a$ is uniformly equal to unity, the above system is analogous to the equations in Refs.~\cite{diebels96, breuer99}. Note that Eq.~\eqref{HypSmall} is not straightforwardly linked to the nonlinear Biot theory by Grinfeld and Norris \cite{grinfeld96} where different inertial terms are proposed.

%%%%%%%%%%%%%%%%%%%%%

\section{Biot's theory}\label{sec:Biot}

%%%%%%%%%%%%%%%%%%%%%%

%+++++++++++++++++++++++++++

\subsection{A linearisation}

%+++++++++++++++++++++++++++

Let us assume that the effective stress in the solid phase satisfies Hooke's law of linear elasticity $\bm{\sigma}^\text{e} = \lambda \operatorname{tr} (\bm{\varepsilon}) \bm{I} + 2 \mu \bm{\varepsilon}$ where $\bm{\varepsilon} = \frac12 \big(\text{grad}\,\bm{u}^\text{s} + \text{grad}^{\sf T}\! \bm{u}^\text{s}\big)$ is the infinitesimal strain tensor, and $\lambda$, $\mu$ are the Lam{\'e} constants.
When we linearise the equations of motion \eqref{HypSmall} about an undeformed static state by neglecting convection terms,  we have
\begin{equation}
	\left\lbrace
	\begin{aligned}
		&\partial_t \bm{\varepsilon} - \tfrac12 \big(\text{grad}\,\bm{v}^\text{s} + \text{grad}^{\sf T}\! \bm{v}^\text{s}\big) = \bm{0}, \\
		&\text{div} \big(n^\text{f}\bm{w} + \bm{v}^\text{s} \big) = 0, \\
		&\rho^\text{f} \partial_t\bm{v}^\text{s} + a \rho^\text{f} \partial_t \bm{w} = -n^\text{f}\text{grad}\, p - \tfrac{(n^\text{f})^2}{k^\text{f}} \bm{w} + \rho^\text{f} \bm{b}^\text{f}, \\
		&\rho \partial_t\bm{v}^\text{s} + \rho^\text{f} \partial_t \bm{w} = \mathrm{div}\, \bm{\sigma}^\text{e} - \text{grad}\, p + \rho\bm{b},
	\end{aligned}\right.
	\label{DiebelsLin}
\end{equation}
where the porosity $n^\text{f} = n^\text{f}_0$, fluid permeability $k^\text{f} = k^\text{f}_0$ and tortuosity $a$ are constant parameters deduced from the values in Table~\ref{tab:Params}.
Eq.~\eqref{DiebelsLin} corresponds exactly to the low-frequency Biot equations with incompressible constituents \cite{bourbie87}, for which Biot's effective-stress coefficient ``$\beta$'' (or ``$\alpha$'' \cite{carcione15}) equals unity and the other Biot parameter ``$M$'' becomes infinite. A more general linearisation about arbitrary pre-deformations in a small-on-large fashion would lead to the acousto-elastic equations, see e.g. Grinfeld and Norris \cite{grinfeld96}. 

%+++++++++++++++++++++++++++

\subsection{Harmonic plane waves}

%+++++++++++++++++++++++++++

We recall the main dispersion characteristics of this theory hereinafter. To do so, harmonic plane-wave motion is assumed by setting the space-time dependence of the unknowns to $\text{e}^{\text{i} (\omega t - k_\omega x)}$, where $\omega$ is the angular frequency, $k_\omega$ is the wave number, and $\text{i} = \sqrt{-1}$ is the imaginary unit.

In the absence of body forces $\bm{b}^\alpha = \bm{0}$, non-trivial solutions to Eq.~\eqref{DiebelsLin} are obtained provided that one of the following dispersion relationships is satisfied:
\begin{equation}
	\begin{aligned}
		(\lambda + 2\mu) \frac{k_\omega ^2}{\omega^2} = \rho^\text{s} + \vartheta\rho^\text{f} - \frac{\text{i}}{\omega k^\text{f}} %\\
		\qquad\text{or}\qquad
		%&
		\mu \frac{k_\omega ^2}{\omega^2} = \rho^\text{s} + \theta\rho^\text{f} + \frac{\rho^\text{f}}{a} \frac{\omega_c^2 - \text{i} \omega_c \omega}{\omega_c^2 + \omega^2} \,  ,
	\end{aligned}
	\label{Dispersion}
\end{equation}
with the coefficients 
\begin{equation}
	\vartheta = \frac{(1-n^\text{f})^2 + a - 1}{(n^\text{f})^2} , \quad
	\theta = \frac{a-1}a \, ,
	\quad
	\omega_c = \frac{n^\text{f}}{a k^\text{f} \rho^\text{fR}} \, .
	\label{Coeffs}
\end{equation}
Note that for the simple mixture model $a \equiv 1$, the above result is the same as that given by De Boer and Liu \cite{deboer94} if fluid compressibility is neglected therein.

In Eq.~\eqref{Dispersion}, the first family of linear waves with elastic modulus $\lambda + 2\mu$ corresponds to longitudinal compression waves (P) resulting from the interaction of both phases, while the second family with elastic modulus $\mu$ corresponds to transverse shear waves (S) mostly supported by the solid skeleton. Recall that although each phase is incompressible, the volume of a particle of mixture can change if the relative quantity of its constituents is modified, i.e. if the porosity is not kept constant (see the comments following Eq.~\eqref{ConstIncomp1}).

\begin{figure}
	\centering
	
	\includegraphics{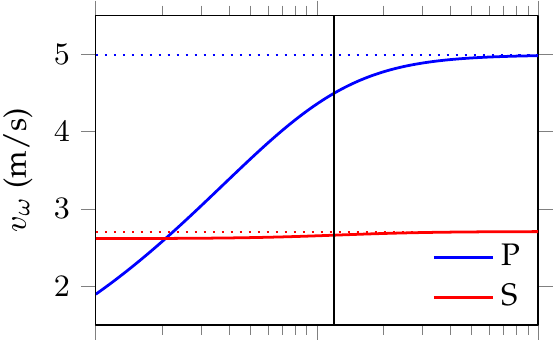}
	
	\hspace{-0.1em}\includegraphics{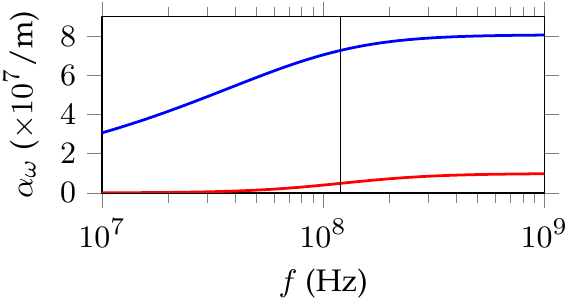}
	
	\caption{Biot's theory:  Dispersion curves (top) and attenuation curves (bottom) for longitudinal compression waves (P) and transverse shear waves (S) deduced from the linearised equations of motion,  with the parameter values of a soft biological tissue saturated by an incompressible liquid in Table~\ref{tab:Params}. The vertical line marks the characteristic frequency $f_c = \omega_c/(2\pi)$ of Eq.~\eqref{Coeffs}, and the horizontal dotted lines are asymptotes. \label{fig:Dispersion}}
	
\end{figure}

Figure~\ref{fig:Dispersion} represents the frequency evolution of the phase velocity $v_\omega = \omega/\text{Re}\, k_\omega$ and of the attenuation coefficient $\alpha_\omega = -\text{Im}\, k_\omega$, for waves propagating towards increasing $x$. The horizontal dotted lines mark the respective high-frequency (or inviscid-fluid) phase velocity limits. As noted by Coussy \cite{coussy04}, ``the undrained situation is recovered (\dots in) the low-frequency range'', where only S-waves propagate.

In the high-frequency range, the analysis reveals that compression and shear waves propagate with strong attenuation resulting from the interaction between the phases.
However, this frequency range is not well-described by the present theory, as viscous dissipation inside the fluid phase becomes preponderant \cite{bourbie87}{\,---\,}for complements and extensions, see the literature on homogenisation theory \cite{auriault09} and enriched continuum models \cite{sciarra07}. Thus, this model is valid in the low-frequency range ${\omega} \ll \omega_c$ where $f_c =\omega_c/(2\pi)$ is a characteristic frequency given in Eq.~\eqref{Coeffs}. In practice, the parameter values of Table~\ref{tab:Params} yield $f_c \approx 120$~MHz (vertical line in Fig.~\ref{fig:Dispersion}), which means that the present set of parameters provides a valid model up to ultrasonic frequencies.

Of course, the value of the critical circular frequency $\omega_c$ plays a crucial role (note in passing that its order of magnitude is related to the characteristic time $\tau_a$ of the Cattaneo-type filtration law \eqref{Cattaneo}). If we use the parameter values of the study by Hosseini-Farid et al. \cite{hosseinifarid20} instead (see also Forte et al. \cite{forte17}), i.e. $n^\text{f}_0 = 0.17$ and $k_0^\text{f} = 1.61 \times 10^{-12}$ m\textsuperscript{2}/(Pa.s), then we find $f_c \approx 4.7$~MHz, which has similar orders of magnitude to the value obtained previously.

Since the propagation characteristics of both waves depend on the linear Biot parameters, acoustic experimental measurements within an adequate frequency range could provide dynamic estimations of those parameters. Ideally, the frequency range of interest should be chosen high enough for the slow P-wave to propagate, but low enough to limit attenuation \cite{bourbie87}.

%%%%%%%%%%%%%%%%%%%%%%%%%

\section{Nonlinear plane waves}\label{sec:nlin}

%%%%%%%%%%%%%%%%%%%%%%%%%

Now we analyse the characteristics of various types of nonlinear wave solutions to Eq.~\eqref{HypSmall} where body forces $\bm{b}^\alpha$ are neglected. We restrict the study to a one-dimensional configuration, assuming invariance along the $y$ and $z$ directions. 

Because the motion does not depend on $y$ and $z$, the deformation gradient and distortion tensors of Eq.~\eqref{F} are of the form
\begin{equation}
	\bm{F} = \begin{bmatrix}
		J & 0 & 0 \\
		-JA_{21} & 1 & 0 \\
		-JA_{31} & 0 & 1
	\end{bmatrix} = \begin{bmatrix}
		J^{-1}& 0 & 0 \\
		A_{21} & 1 & 0 \\
		A_{31} & 0 & 1
	\end{bmatrix}^{-1} = \bm{A}^{-1} ,
	\label{F1D}
\end{equation}
where $J = \det \bm{F}$ denotes the volume dilatation given by $J = F_{11} = (A_{11})^{-1}$. 
Therefore, the motion includes possibly a volume-changing compressive deformation ($11$-component) and a volume-preserving shear deformation ($21$- and $31$-components).

With the above invariances in mind, the system \eqref{HypSmall} can be rewriten as a \emph{first-order quasi-linear system} of partial differential equations,
\begin{equation}
	{\bf M}^t({\bf q})\, \partial_t {\bf q} + {\bf M}^x({\bf q})\, \partial_x {\bf q} = {\bf R}({\bf q}),
	\label{Hyp}
\end{equation}
for the vector ${\bf q} = [A_{11}, A_{21}, A_{31}, n^\text{f}, v^\text{s}_1, v^\text{s}_2, v^\text{s}_3, w_1, w_2, w_3, p]^{\sf T}\!$, where the coefficients of ${\bf M}^\nu$ for $\nu \in \lbrace t, x\rbrace$ and $\bf R$ are specified in Appendix~\ref{app:Mat}. Note in passing that these arrays do not depend on $v^\text{s}_2$, $v^\text{s}_3$ and $p$. Moreover, they are not symmetric, and the matrix ${\bf M}^t$ is not invertible.

%+++++++++++++++++++++++++++

\subsection{Smooth travelling waves}

%+++++++++++++++++++++++++++

We consider plane wave solutions propagating with constant speed $c$, such that the field variables are smooth functions of $\xi = x-c t$ only.
Hereinafter, primes $'$ denote differentiation with respect to $\xi$, so that  $\partial_x = (\cdot)'$ and $\partial_t = -c\, (\cdot)'$ according to the chain rule. 
Hence, our system \eqref{Hyp} reduces to the ordinary differential system
\begin{equation}
	{\bf M}({\bf q})\, {\bf q}' = {\bf R}({\bf q}),
	\label{Diff}
\end{equation}
with the matrix-valued function ${\bf M} = {\bf M}^x - c\, {\bf M}^t$ of the vector ${\bf q}$.

Bounded solutions of Eq.~\eqref{Diff} that connect two equilibrium states are called \emph{travelling waves}. One may be able to derive such solutions in the case where ${\bf M}$ is invertible, by rewriting Eq.~\eqref{Diff} as an autonomous dynamical system. However, this is not as straightforward in practice. In fact, the bad conditioning of the matrix ${\bf M}$ makes the classical analysis difficult. Attempts to exhibit such solutions numerically have been unsuccessful up to now, suggesting that travelling waves may not propagate in such a material. Viscous dissipation, compressibility or compaction might be needed for this peculiar nonlinear phenomenon to emerge, see e.g. Refs.~\cite{depascalis19, jordan17, connolly15}.

\subsection{Characteristic wave speeds}

Let us consider particular wave solutions for which the matrix ${\bf M}({\bf q})$ is singular.
To do so, let us focus on the homogeneous system ${\bf M}({\bf q})\, {\bf q}' = {\bf 0}$ by setting ${\bf R}$ equal to zero. According to Eq.~\eqref{Hyp1D} of the Appendix, this amounts to assuming inviscid flow for which $k^\text{f} \to +\infty$. Non-trivial solutions for $\bf q$ can be obtained if $\det {\bf M}({\bf q})$ vanishes, restricting the value of the wave speed $c$ to one of the generalised eigenvalues of ${\bf M}^x$ and ${\bf M}^t$.

In general, it is a difficult task to compute these characteristic wave velocities analytically. If the material has a neo-Hookean behaviour ($\beta=0$), then we find that the wave speed $c$ equals one of the following values:
\begin{equation}
	\begin{aligned}
		&c^\pm_P = v^\text{s}_1 + \frac{\frac12(2\vartheta - \vartheta^*) \rho^\text{f} w_1 %}{ \rho^\text{s} + \vartheta\rho^\text{f} } \\ &
			\pm %\frac{
			\sqrt{ (\rho^\text{s} + \vartheta\rho^\text{f}) A_{11}Q_{11} + (\frac12\vartheta^* \rho^\text{f} w_1)^2 + (\vartheta^*-\vartheta) \rho^\text{s}\rho^\text{f} w_1^2 } }{ \rho^\text{s} + \vartheta\rho^\text{f} } ,\\[2pt]
		&c^\pm_S = v^\text{s}_1 + \frac{\frac12\theta \rho^\text{f} w_1 \pm \sqrt{\left(\rho^\text{s} + \theta \rho^\text{f}\right) Q_{22} + (\frac12\theta \rho^\text{f} w_1)^2}}{\rho^\text{s} + \theta \rho^\text{f}} , \qquad %\\ &
		c^\text{f}\, = v^\text{f}_1 , \qquad c^\text{s}\, = v^\text{s}_1 ,
	\end{aligned}
	\label{Speed}
\end{equation}
where $\vartheta^* = (a-1)/n^\text{f} \geq 0$ and the coefficients $\vartheta$, $\theta \geq 0$ defined in Eq.~\eqref{Coeffs} are functions of the porosity $n^\text{f}$, itself function of the compression strain $A_{11}$ according to the continuity equation \eqref{ContinuityEIncomp}. The coefficients $Q_{ij} = -\partial {\sigma}^\text{e}_{i1}/\partial A_{j1}$ specified in Appendix~\ref{app:Mat} are functions of the skeleton's deformation. Since the material's behaviour is assumed neo-Hookean ($\beta=0$), these coefficients satisfy $Q_{22} = Q_{33}$, and the four coefficients $Q_{12}$, $Q_{13}$, $Q_{23}$, $Q_{32}$ are equal to zero. Note in passing that the speed of P-waves has a nonlinear expression with respect to the compression strain $A_{11}$, while the speed of S-waves is independent on the shear strains $A_{21}$, $A_{31}$.

Acoustic waves propagate, i.e. hyperbolicity is ensured, if the sound speeds in Eq.~\eqref{Speed} are real. For this purpose, the radical's argument in the expression of $c^\pm_P$ and $c^\pm_S$ must be non-negative. In both cases, one notes that this quantity is of the form $\mathfrak{a} + \mathfrak{b} w_1^2$. The propagation condition $\mathfrak{a} + \mathfrak{b} w_1^2 \geq 0$ can be simplified if the coefficient $\mathfrak{b}$ is non-negative, in which case imposing $\mathfrak{a} \geq 0$ will be sufficient to ensure hyperbolicity.

While the analysis of hyperbolicity is straightforward for shear waves with speed $c^\pm_S$ in which case $\mathfrak{b}$ is non-negative, the analysis is less obvious for compression waves with speed $c^\pm_P$. In the case of simple mixtures ($a\equiv 1$), the coefficient $\mathfrak{b}$ in the expression of $c^\pm_P$ has the same sign as $(n^\text{f}-1)$; hence, it is always negative. If Berryman's formula \eqref{Tortuosity} is used instead ($a\not\equiv 1$), this coefficient has the same sign as $(n^\text{f}-1) (n^\text{f}+\frac12) + \epsilon$, where the constant $\epsilon = \frac1{16} \rho^\text{fR}/\rho^\text{sR}$ depends on the ratio of the reference densities. In the present low-porosity material with $\rho^\text{fR} \simeq \rho^\text{sR}$, see values in Table~\ref{tab:Params}, the coefficient $\mathfrak{b}$ in the expression of $c^\pm_P$ is negative. Thus, the propagation of compression waves requires that the seepage velocity has a moderate amplitude $|w_1|$ along the direction of propagation.

\begin{condition}\label{Hyperbo}
	{\bfseries\upshape (Hyperbolicity)}
	From the expression of the characteristic wave speeds in Eq.~\eqref{Speed}, a sufficient condition of hyperbolicity reads
	\begin{equation}
		A_{11}Q_{11} = -A_{11}\frac{\partial {\sigma}^\textup{e}_{11}}{\partial A_{11}} \geq -\mathfrak{b} w_1^2   , \qquad Q_{22} = -\frac{\partial {\sigma}^\textup{e}_{21}}{\partial A_{21}} \geq 0  
		\label{CondHyperbo}
	\end{equation}
	with $\mathfrak{b}$ deduced from the expression of $c^\pm_P$. Under this condition, any plane wave propagates with finite speed within the biphasic neo-Hookean model \eqref{F1D}-\eqref{Hyp} where $\beta = 0$.
\end{condition}

\begin{figure}
	\centering
	\includegraphics{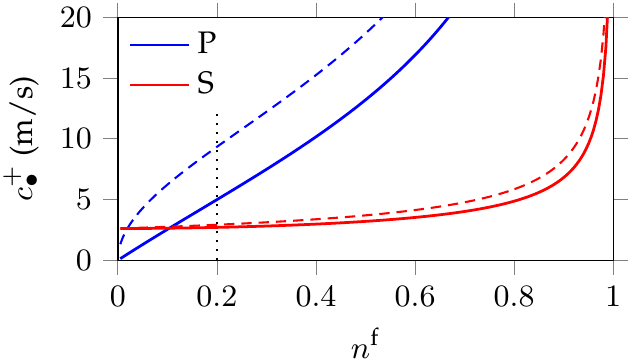}
	
	\caption{Evolution of the characteristic speeds $c^+_P$, $c^+_S$ of Eq.~\eqref{Speed} in terms of the porosity in a static undeformed configuration. Porosity-dependent tortuosity ($a \not\equiv 1$, solid lines) is compared with the case of a simple mixture ($a \equiv 1$, dashed lines) using the parameters of Table~\ref{tab:Params}. \label{fig:SpeedNL}}
\end{figure}

Figure~\ref{fig:SpeedNL} displays the evolution of the above characteristic wave speeds \eqref{Speed} with porosity, at zero velocity and no strain{\,---\,}in other words, a static undeformed state of the form ${\bf q} = [1, 0, 0, n^\text{f}, 0, 0, 0, 0, 0, 0, p]^{\sf T}\!$ is considered. Therefore, the speeds of sound become
\begin{equation}
	c^\pm_P = \pm \sqrt\frac{\lambda + 2\mu}{ \rho^\text{s} + \vartheta\rho^\text{f} } ,\qquad
	c^\pm_S = \pm \sqrt\frac{\mu}{\rho^\text{s} + \theta \rho^\text{f}} , \qquad c^\text{f}\, = 0 , \qquad c^\text{s}\, = 0 ,
	\label{SpeedEq}
\end{equation}
with the coefficients of Eq.~\eqref{Coeffs}.
These wave speeds are the same as the phase velocities deduced from Biot's theory \eqref{DiebelsLin} in the inviscid fluid limit $k^\text{f} \to +\infty$, or equivalently in the high-frequency limit (horizontal dotted lines in Fig.~\ref{fig:Dispersion}). In the variable tortuosity case \eqref{Tortuosity}, the porosity $n^\text{f}_0 = 0.2$ of Table~\ref{tab:Params} yields the values $4.99$~m/s and $2.71$~m/s for $c^+_P$ and $c^+_S$ (vertical dotted line in Fig.~\ref{fig:SpeedNL}).

Using Berryman's formula \eqref{Tortuosity}, the coefficients in Eq.~\eqref{SpeedEq} satisfy $\vartheta \to +\infty$ and $\theta \to 1$ at zero porosity. At unit porosity, they satisfy $\vartheta$, $\theta \to 0$. Thus, as shown in the figure (solid lines), compression waves do not propagate at zero porosity, and both waves do not propagate at unit porosity. The first remark relates to the fact that the monophasic solid limit $n^\text{f}\to 0$ is an incompressible solid in which shear waves propagate, but not poroelastic compression waves. The second remark expresses the fact that the monophasic fluid limit $n^\text{f}\to 1$ does not support shear stresses nor poroelastic compression.

Fig.~\ref{fig:SpeedNL} compares the evolution obtained for variable tortuosity \eqref{Tortuosity} to that of the simple mixture model where the tortuosity coefficient $a \equiv 1$ is not porosity-dependent. One notes that the tortuosity effect has a significant influence on the P-wave velocity at low porosities, while the shear-wave velocity does not seem to be significantly affected by this feature, see also Wilmanski \cite{wilmanski05} where comparisons between Biot's theory and the simple mixture model are proposed. In a different context, the fact that the tortuosity factor has a major influence on the speed of the slow P-wave is a well-known feature \cite{johnson82}.

\subsection{Perturbation approach}

Let us investigate the influence of the Yeoh parameter $\beta$ on the characteristic speeds by using a perturbation method \cite{nayfeh00}. For this purpose, we introduce pairs of left and right generalised eigenvectors ${\bf l}$, ${\bf r}$ deduced from the condition ${\bf M}({\bf q}) \, {\bf q}' = {\bf 0}$. The vectors $\bf r$ form a basis of the right null space of ${\bf M}$, while the vectors $\bf l$ form a basis of the left null space of ${\bf M}$, i.e. they belong to the right null space of ${\bf M}^{\sf T}$.

As shown in the matrices' expression (Appendix~\ref{app:Mat}), the matrix ${\bf M}^x$ is linear in $\beta$, but ${\bf M}^t$ does not depend on $\beta$. This can be rewritten as a perturbation of the form ${\bf M}^x = {\bf M}^{x0} + \beta {\bf M}^{x1}$, where the zeroth-order matrix ${\bf M}^{x0}$ corresponds to the neo-Hookean case discussed in the previous section. Thus, we seek generalised eigenvalues and eigenvectors as power series of $\beta$:
\begin{equation}
	c = c^0 + \beta c^1 + \dots , \qquad {\bf l} = {\bf l}^0 + \beta {\bf l}^1 + \dots , \qquad {\bf r} = {\bf r}^0 + \beta {\bf r}^1 + \dots
	\label{Perturbation}
\end{equation}
where the zeroth-order quantities $c^0$, ${\bf l}^0$, ${\bf r}^0$ correspond to the case of neo-Hookean behaviour ($\beta=0$).
Injecting this Ansatz in the generalised eigenvalue problems ${\bf M}{\bf r} = {\bf 0}$ and ${\bf l}^{\sf T}{\bf M} = {\bf 0}$ leads to the conditions
\begin{equation}
	\begin{aligned}
		\text{order 0:} \qquad &{\bf M}^0{\bf r}^0 = {\bf 0} \, , & \qquad &{\bf l}^{0 {\sf T}} {\bf M}^0 = {\bf 0} \, ,\\
		\text{order 1:} \qquad &{\bf M}^1{\bf r}^0 + {\bf M}^0{\bf r}^1 = {\bf 0} \, , & \qquad &{\bf l}^{0 {\sf T}}{\bf M}^1 + {\bf l}^{1 {\sf T}}{\bf M}^0 = {\bf 0} \, ,
	\end{aligned}
	\label{PerturbationCond}
\end{equation}
with ${\bf M}^p = {\bf M}^{xp} - c^p {\bf M}^{t}$ and $p\in \lbrace 0,1\rbrace$, at zeroth order and first order in $\beta$.

Now, we left-multiply the vector ${\bf M}^1{\bf r}^0 + {\bf M}^0{\bf r}^1$ by the vector ${\bf l}^{0 {\sf T}}$. Thus, the zeroth-order identity ${\bf l}^{0 {\sf T}} {\bf M}^0 = {\bf 0}$ leads to the following approximate expression of the Yeoh characteristic speeds
\begin{equation}
	c \simeq c^0 + \frac{{\bf l}^{0 {\sf T}} (\beta {\bf M}^{x1}) {\bf r}^{0}}{{\bf l}^{0 {\sf T}} {\bf M}^{t} {\bf r}^{0}} 
	\label{SpeedYeoh}
\end{equation}
at first order in $\beta$. One observes that the increment of the speed of sound is linear with respect to the (presumably small) perturbation $\beta {\bf M}^{x1}$ of the matrix ${\bf M}^{x}$. In practice, the pairs of vectors ${\bf l}^0$, ${\bf r}^0$ deduced from previous section are normalised in such a way that ${\bf l}^{0 {\sf T}} {\bf M}^{t} {\bf r}^0 = 1$, which greatly simplifies Eq.~\eqref{SpeedYeoh}.

\paragraph{Compression waves.} Let us go back to the zeroth-order neo-Hookean case. Using a computer algebra system, one pair of vectors ${\bf l}^0$, ${\bf r}^0$ is deduced from ${\bf M}({\bf q})$ by solving the generalised eigenvalue problem corresponding to the characteristic speed $c^0_P = c^+_P$ of Eq.~\eqref{Speed}. The components of these vectors lead to the perturbation \eqref{SpeedYeoh}
\begin{equation}
	c_P^+ \simeq c^0_P + \frac{1}{2\sqrt{\lambda+2\mu}}\frac{Q_{11}^1}{\sqrt{\rho^\text{s} + \vartheta \rho^\text{f}}} , \qquad Q_{11}^1 = \mu\beta \frac{3-A_{11}^2+3 A_{21}^2+3 A_{31}^2}{A_{11}^4}
	\label{PerturbComp}
\end{equation}
of the speed of compression waves. The quantity $Q_{11}^1$ denotes the first-order increment of the coefficient $Q_{11}$ given in Appendix~\ref{app:Mat}. Note that the speed of sound is no longer exclusively function of volume-changing strain $A_{11}$, and that the above perturbation has a non-zero value in undeformed state.

\paragraph{Shear waves.} Solving the generalised eigenvalue problem corresponding to the characteristic speed $c^0_S = c^+_S$ of Eq.~\eqref{Speed} leads to the perturbation \eqref{SpeedYeoh}
\begin{equation}
	c_S^+ \simeq c^0_S + \frac{1}{2\sqrt{\mu}}\frac{Q_{22}^1}{\sqrt{\rho^\text{s} + \theta \rho^\text{f}}} , \qquad Q_{22}^1 = \mu\beta \frac{1-A_{11}^2+3 A_{21}^2+A_{31}^2}{A_{11}^3}
	\label{PerturbShear}
\end{equation}
of the shear wave speed with polarisation along $y$, where the coefficient $Q_{22}^1$ is deduced from the Appendix~\ref{app:Mat}. One notes that the speed of sound is no longer independent on the shear deformation $A_{21}$, $A_{31}$ and that this dependency is quadratic, which is coherent with related studies \cite{tripathi19}. Here, the sound speed in an undeformed state \eqref{SpeedEq} is unchanged. An expression similar to Eq.~\eqref{PerturbShear} is found for shear waves polarised along $z$, where the increment $Q_{22}^1$ needs to be replaced by a coefficient $Q_{33}^1$ obtained in a similar fashion from the expressions in the Appendix.

\bigskip

Figure~\ref{fig:Perturb} illustrates the validity of the above perturbations. Fig.~\ref{fig:Perturb}a compares the perturbation \eqref{PerturbComp} of the Yeoh P-wave speed (blue dashed line) with the same value obtained by numerical resolution of the generalised eigenvalue problem of ${\bf M}^x$ and ${\bf M}^t$ (blue solid line). The value of the perturbation parameter $\beta = 2.2$ is taken from Table~\ref{tab:Params}, as well as the value of other parameters. Here, a static state under pure dilatation is considered, i.e. the shear strain components $A_{21}$, $A_{31}$ are set to zero while $A_{11}$ is varied. Thus, the porosity $n^\text{f} = 1 - 0.8\, A_{11}$ is not constant. The agreement between both curves is very good in the vicinity of the static undeformed state $A_{11}\simeq 1$. Similary, Fig.~\ref{fig:Perturb}b illustrates the effect of the perturbation \eqref{PerturbShear} on the shear wave speed. Here, a static state under simple shear is considered, i.e. $A_{11}=1$ and $A_{31}=0$ are imposed while $A_{21}$ is varied, and the porosity $n^\text{f} = 0.2$ is constant.

\begin{figure}
	\begin{minipage}{0.49\textwidth}
		\centering
		(a)
		
		\includegraphics{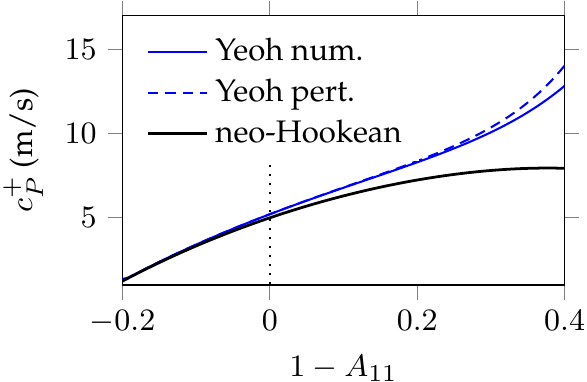}
	\end{minipage}
	\begin{minipage}{0.49\textwidth}
		\centering
		(b)
		
		\includegraphics{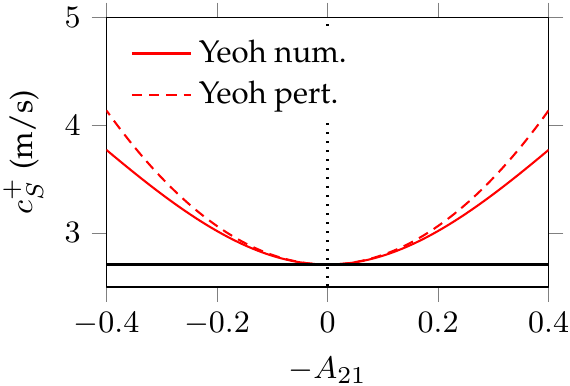}
	\end{minipage}

	\caption{Perturbation approach. Evolution of the Yeoh characteristic speeds \eqref{SpeedYeoh} with respect to the strain components using the reference parameter values of Table~\ref{tab:Params}; (a) P-wave velocity under purely volume-changing deformations, (b) S-wave velocity under purely isochoric simple shear deformations. The black lines mark the neo-Hookean case \eqref{Speed}. \label{fig:Perturb}}
\end{figure}

In both figures, the black solid line corresponds to the neo-Hookean case \eqref{Speed} where $\beta = 0$, and the vertical dotted line marks the undeformed state. In Fig.~\ref{fig:Perturb}b, one observes that the situation is almost symmetric with respect to the undeformed state, and that the neo-Hookean case yields a strain-independent shear wave speed. The picture is different in Fig.~\ref{fig:Perturb}a, where the curves are not symmetric with respect to the undeformed state, and where the neo-Hookean model yields already strain-dependent sound velocities. These observations suggest that the nonlinearity of poroelastic P-wave propagation is of very different nature to that of shear wave propagation.

\subsection{Acceleration waves}

Similarly to Refs.~\cite{deboer95, wilmanski05, ciarletta18}, let us analyse the speed and evolution of \emph{acceleration waves}. For such wave solutions, the primary field $\bf q$ is continuous across the surface $\xi(\bm{x}, t) = 0$ with $\xi = x - s(t)$, but its normal derivative $\partial_\xi {\bf q}$ may be discontinuous. Typically, such solutions represent situations in which the field variables experience a brutal change of slope; for instance, an initial-value problem with piecewise linear initial data.

As proposed by M{\"u}ller and Ruggeri \cite{ruggeri98}, we assume that the wave propagates into a domain where the primary field ${\bf q}$ is a constant equilibrium state $\bar{\bf q}$ of Eq.~\eqref{Hyp} for which the seepage velocity $\bar{\bm w}$ equals zero{\,---\,}more general cases are discussed in the literature \cite{ruggeri98}. The jumps $\llbracket \cdot \rrbracket$ of the partial derivatives across the moving surface are related to those of the normal derivative $\partial_\xi {\bf q}$ according to $\llbracket \partial_t {\bf q} \rrbracket = -c\, \llbracket \partial_\xi {\bf q} \rrbracket$ and $\llbracket \partial_x {\bf q} \rrbracket = \llbracket \partial_\xi {\bf q} \rrbracket$,
where the speed satisfies $c = \partial_t s$. Therefore, by computing the jump of Eq.~\eqref{Hyp} and using the continuity requirement $\llbracket {\bf q} \rrbracket = {\bf 0}$, we find
\begin{equation}
	{\bf M}(\bar{\bf q})\, \llbracket \partial_\xi {\bf q} \rrbracket = {\bf 0}
	\label{Acceleration}
\end{equation}
along the wavefront.
Non-trivial solutions to Eq.~\eqref{Acceleration} are found if ${\bf M}(\bar{\bf q})$ is singular, i.e. if $c$ equals one of the characteristic velocities of Eq.~\eqref{Speed} evaluated at $\bar{\bf q}$. Then, the jump vector $\llbracket \partial_\xi {\bf q} \rrbracket$ belongs to the kernel of ${\bf M}$, or equivalently, to the corresponding generalised eigenspace of ${\bf M}^x$ and ${\bf M}^t$. In other words, we may write that $\llbracket \partial_\xi {\bf q} \rrbracket = \Pi\, {\bf r}$ is proportional to a basis vector $\bf r$ of the right null space of ${\bf M}(\bar{\bf q})$. If $\bf r$ is scaled in such a way that it has same dimension as $\bf q$ componentwise, then the \emph{wave amplitude} $\Pi$ is expressed in $\text{m}^{-1}$.

Now we derive Bernoulli's evolution equation satisfied by the wave amplitude following Sec.~8.4 of Ref.~\cite{ruggeri98}{\,---\,}a similar result was obtained by Ciarletta et al. \cite{ciarletta18} for general Biot-like models. For this purpose, we consider a vector ${\bf l}$ belonging to the left null space of ${\bf M}(\bar{\bf q})$, and such that ${\bf l}^{\sf T} {\bf M}^t\, {\bf r} = 1$. As shown in the literature \cite{ruggeri98}, a Bernoulli differential equation is obtained
\begin{equation}
	\frac{\text d}{\text d t} \Pi+ \Omega_1 \Pi + \Omega_2 \Pi^2 = 0 \, ,
	\label{Bernoulli}
\end{equation}
where ${\text d}/{\text d t}$ denotes the directional derivative $\partial_t + c\,\partial_x$ along the curve that follows the position of the surface. A well-known analytical solution to Eq.~\eqref{Bernoulli} yields the time-evolution,
\begin{equation}
	\Pi(t) = \frac{\Pi(0)\, \text{e}^{-\Omega_1 t}}{1 + \Pi(0) \frac{\Omega_2}{\Omega_1}(1 - \text{e}^{-\Omega_1 t})},
	\label{BernoulliSol}
\end{equation}
of the jump amplitude as the wave propagates, in terms of the coefficients
\begin{equation}
	\Omega_1 = -{\bf l}^{\sf T} \frac{\partial {\bf R}}{\partial {\bf q}}\, {\bf r}
	\, , \qquad 
	\Omega_2 = \left(\frac{\partial c}{\partial {\bf q}}\right)^{\!{\sf T}} {\bf r},
	\label{BernoulliCoeff}
\end{equation}
evaluated at the constant equilibrium state $\bar{\bf q}$.

Assuming positive coefficients $\Omega_1$, $\Omega_2$ in the expression \eqref{BernoulliSol} of the wave amplitude, we observe that the denominator vanishes at some finite time,
\begin{equation}
	t_s = -\frac{1}{\Omega_1} \ln \left(1 + \frac1{\Pi(0)}\frac{\Omega_1}{\Omega_2}\right) > 0,
	\label{CriticalTime}
\end{equation}
if the initial jump $\Pi(0)$ is smaller than $-\Omega_1/\Omega_2$.
Conversely, such acceleration wave solutions are stable for positive times under the condition $\Pi(0) > -\Omega_1/\Omega_2$.

Note that $\Omega_1$ vanishes in the case of inviscid flow $k^\text{f} \to +\infty$. Since $\Omega_1$ depends on ${\bf R}$, it accounts for attenuation. As can be seen from Eq.~\eqref{BernoulliSol}, the constant $\Omega_1$ is responsible for the decay of the jump amplitude, and therefore provides a smoothing effect on wave solutions. The constant $\Omega_2$ vanishes when the characteristic speed $c$ corresponds to a \emph{linearly degenerate} eigenspace \cite{godlewski96}. Thus, this constant expresses the nonlinearity of wave propagation, and therefore may yield a competing wavefront steepening effect.

\paragraph{Compression waves.}

Assume that $\bar{\bf q} = [1, 0, 0, n^\text{f}_0, 0, 0, 0, 0, 0, 0, \bar p]^{\sf T}\!$ corresponds to a motionless undeformed equilibrium state, and that the material's behaviour is neo-Hookean ($\beta=0$). Using a computer algebra system, one pair of vectors ${\bf l}$, ${\bf r}$ is deduced from ${\bf M}(\bar{\bf q})$ by solving the generalised eigenvalue problem corresponding to the characteristic speed $c = c^+_P \approx 4.99$~m/s, which is the value displayed in Fig.~\ref{fig:SpeedNL} at the vertical dotted line. The components of $\bf r$ lead to the particular relationship
\begin{equation}
	\llbracket \partial_\xi v_1^\text{s} \rrbracket + n^\text{f}\, \llbracket \partial_\xi w_1 \rrbracket = 0,
\end{equation}
between the acceleration jumps,
showing that acceleration P-waves propagate both in the fluid and the solid phase. This relationship is the same as that found by De Boer and Liu \cite{deboer95} in the case of linearly-elastic simple mixtures. The jump amplitude is governed by the Bernoulli equation \eqref{Bernoulli} with the coefficients \eqref{BernoulliCoeff}
\begin{equation}
	\begin{array}{c}
		\displaystyle
		\Omega_1 = \frac{(n^\text{f})^2/{k^\text{f}} }{2(n^\text{f})^2 (\rho^\text{s} + \vartheta\rho^\text{f}) } \, , \quad
		\\[1em]
		\displaystyle
		\Omega_2 = \frac{ \big(  (\mu+\frac32\lambda) (n^\text{f})^3 - (4\mu + 3\lambda) (n^\text{f})^2 + \mu n^\text{f} + \lambda + 2 \mu \big) \frac{\rho^\text{f}}{(n^\text{f})^4} %}{ \sqrt{\lambda + 2\mu} \, (\rho^\text{s} + \vartheta\rho^\text{f})^{3/2} } %\\&\quad
			+ %\frac{
			(\mu+\frac32\lambda) \frac{\rho^\text{s}}{n^\text{f} - 1} }{ \sqrt{\lambda + 2\mu} \, (\rho^\text{s} + \vartheta\rho^\text{f})^{3/2} }
	\end{array}
	\label{CoeffsP}
\end{equation}
evaluated at $\bar{\bf q}$. While the coefficient $\Omega_1$ depends on tortuosity through $\vartheta$, the explicit dependence of $\Omega_2$ on tortuosity is no longer apparent in Eq.~\eqref{CoeffsP} where Berryman's formula \eqref{Tortuosity} was used. The values of Table~\ref{tab:Params} yield the characteristic distance of decay $c^+_P/\Omega_1 \approx 12.4$~nm, as well as the critical wave amplitude $-\Omega_1/\Omega_2 \approx -1.02 \times 10^{7}$~$\text{m}^{-1}$. The expression of $\Omega_1$ in Eq.~\eqref{CoeffsP} is the same as that proposed by De Boer and Liu \cite{deboer96b} for the linearised simple mixture. While longitudinal acceleration waves decay exponentially in the linear case $\Omega_2 = 0$, this property is no longer true at large amplitudes in the present nonlinear case.

\paragraph{Shear waves.}

Again, $\bar{\bf q} = [1, 0, 0, n^\text{f}_0, 0, 0, 0, 0, 0, 0, \bar p]^{\sf T}\!$ corresponds to a motionless undeformed equilibrium state, and the material's behaviour is assumed neo-Hookean ($\beta=0$).
For the characteristic speed $c = c^+_S \approx 2.71$~m/s,
two distinct pairs of vectors ${\bf l}$, ${\bf r}$ are found, corresponding to shear waves polarised along $y$ or $z$. For both polarisations ($i \in \lbrace 2, 3\rbrace$), we find that the acceleration jumps are linked through
\begin{equation}
	\llbracket \partial_\xi v_i^\text{s} \rrbracket + a\, \llbracket \partial_\xi w_i \rrbracket = 0 \, ,
\end{equation}
where the tortuosity coefficient $a$ is deduced from the porosity in the state $\bar{\bf q}$.
This relationship suggests that acceleration S-waves propagate both in the fluid and the solid phase if $a \neq 1$. If $a = 1$, these waves propagate only in the solid phase, as shown by De Boer and Liu \cite{deboer95} for linearly-elastic simple mixtures. The jump amplitude is governed by Bernoulli's equation \eqref{Bernoulli} with the coefficients \eqref{BernoulliCoeff}
\begin{equation}
	\Omega_1 = \frac{{(n^\text{f})^2}/{k^\text{f}} }{2a^2 (\rho^\text{s} + \theta\rho^\text{f}) }
	\, ,\qquad
	\Omega_2 = 0
	\label{CoeffsS}
\end{equation}
evaluated at $\bar{\bf q}$. Therefore, transverse acceleration waves decay exponentially, consistently with the study by De Boer and Liu \cite{deboer96b} where $a \equiv 1$. The characteristic distance of decay deduced from Table~\ref{tab:Params} is $c^+_S/\Omega_1 \approx 101$~nm. Note that the order of magnitude of this characteristic distance relates to that of the high-frequency attenuation distance $\alpha_\omega^{-1}$ deduced from the dispersion analysis (see Fig.~\ref{fig:Dispersion}).

\bigskip

\begin{figure}
	\begin{minipage}{0.49\textwidth}
		\centering
		(a)
		
		\includegraphics{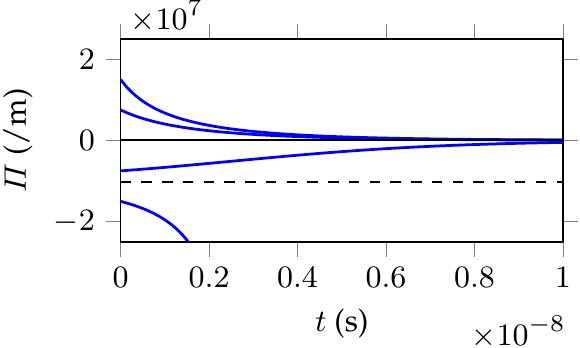}
	\end{minipage}
	\begin{minipage}{0.49\textwidth}
		\centering
		(b)
		
		\includegraphics{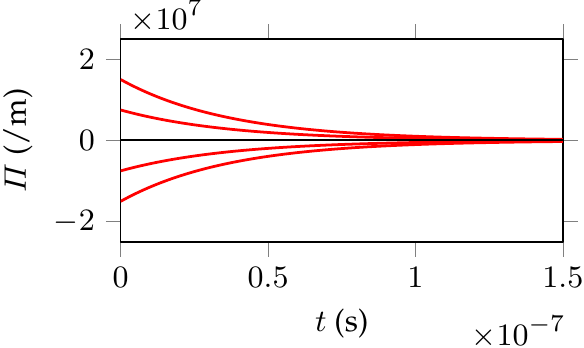}
	\end{minipage}
	
	\caption{Evolution of the amplitude of nonlinear acceleration waves deduced from Eq.~\eqref{BernoulliSol} for initial amplitudes $\Pi(0)$ ranging from $-1.5$ to $1.5 \times 10^{7}$ /m. (a) Longitudinal waves and (b) transverse waves. \label{fig:Evol}}
\end{figure}

Figure~\ref{fig:Evol} displays the time-evolution of the amplitude $\Pi$ deduced from Eq.~\eqref{BernoulliSol}, for nonlinear poroelastic acceleration P-waves and S-waves propagating in a neo-Hookean material ($\beta=0$). While shear wave amplitudes decay exponentially (smoothing effect), compression waves are subject to a critical amplitude $-\Omega_1/\Omega_2$ deduced from Eq.~\eqref{CoeffsP} (horizontal dashed line in Fig.~\ref{fig:Evol}a), below which the solution becomes infinite in finite time. As stated in M{\"u}ller and Ruggeri \cite{ruggeri98} p.~183, ``if the initial discontinuity in the derivatives is too strong, it cannot be damped; instead it grows to infinity and thus the acceleration wave develops into a shock wave''. Beyond the critical amplitude, nonlinearity overpowers attenuation effects, leading to the formation of shock waves.

Of course, the addition of Yeoh behaviour with $\beta > 0$ modifies the picture slightly (see Fig.~\ref{fig:Perturb}, and the modified sound velocities in Eq.~\eqref{SpeedYeoh}), since the shear sound velocities are no longer independent on the shear deformation. Therefore, the linear degeneracy property for shear waves will potentially be lost. Nevertheless, since the shear wave speed is quadratic in the shear strains, the values found for $\Omega_1$, $\Omega_2$ should not be greatly affected by this modification, at least about an undeformed state where the sound velocity is nearly constant. In a small deformation range, the significant difference of magnitude for the coefficients $\Omega_1$, $\Omega_2$ in compression and shear waves will remain a predominant feature of the material. Therefore, shock waves will still develop more easily in compression than in shear, while poroelastic P-waves are subject to faster smoothing than S-waves due to stronger attenuation.

\section{Conclusion}\label{sec:Conclu}

A mixture-theoretic Biot model for large deformations in incompressible media has been presented, in view of future biomechanical applications. Here, saturated Yeoh-type porous solids were considered. The main features are the existence of shear waves and slow compression waves, which linear dispersive properties follow from the Biot theory. The computation of plane-wave solutions with discontinuous gradients shows that shear jumps decay exponentially (in a similar fashion to the linear theory), while the compression jumps are governed by a nonlinear Bernoulli equation. Thus, in the neo-Hookean limit, large compressive jumps can lead to the formation of shocks, which is not the case of shear jumps.

These results can be used for the validation of numerical methods \cite{diebels96,heider12}. Moreover, the modelling framework and the methodology are applicable to other fields, for instance where wave propagation problems in compressible or triphasic mixtures arise. As discussed above, shock waves may emerge. In this regard, a first difficulty lies in the quasi-linear (non-conservative) form of the equations of motion, for which the definition of shock wave solutions is not straightforward \cite{ruggeri98}. While this problem can be circumvented in the case of simple mixtures $a \equiv 1$, the derivation of a conservative form is less obvious in the case of porosity-dependent tortuosity coefficients $a \not\equiv 1$. Nevertheless, shock wave solutions can still be investigated numerically. A possible strategy would be to rely on shock-capturing finite volume methods \cite{godlewski96}, e.g. based on an `artificial compressibility' approach to account for the saturation constraint.

As far as the present problem is concerned, several improvements need to be mentioned. First, one should be aware that the generality of the results is limited by the constitutive assumptions, but that the same approach could be used for variations of this model. Second, the use of poroelasticity in applications requires the experimental determination of relevant model parameters (Table~\ref{tab:Params}). In practice, the brain mechanics literature suffers from a lack of experimental data in dynamic configurations, that would be representative of head trauma configurations. Lastly, several modelling refinements could be introduced in potential fine tuning steps, such as viscoelastic behaviour \cite{hosseinifarid20,comellas20}, non-Darcy flow \cite{markert07} or objective derivatives \cite{wilmanski05} to name a few. A conclusive experimental or computational assessment of multiphasic effects in TBI is still needed.

\subsection*{Acknowledgements}
This work was supported by the Irish Research Council [project ID GOIPD/2019/328].
The author is grateful to Michel Destrade (NUI Galway) for fruitful discussions, careful reading and support.

%\disclaimer{}

%%%%%%%%%% Insert bibliography here %%%%%%%%%%%%%%

%\vskip2pc

%\bibliographystyle{RS}
\bibliography{Biblio}

\begin{thebibliography}{10}
\expandafter\ifx\csname url\endcsname\relax
  \def\url#1{\texttt{#1}}\fi
\expandafter\ifx\csname urlprefix\endcsname\relax\def\urlprefix{URL }\fi
\expandafter\ifx\csname href\endcsname\relax
  \def\href#1#2{#2} \def\path#1{#1}\fi

\bibitem{deboer96a}
R.~De~Boer, {Highlights in the historical development of the porous media
  theory: Toward a consistent macroscopic theory}, Appl. Mech. Rev. 49~(4)
  (1996) 201--262.
\newblock \href {https://doi.org/10.1115/1.3101926}
  {\path{doi:10.1115/1.3101926}}.

\bibitem{ateshian17}
G.~A. Ateshian, Mixture theory for modeling biological tissues: Illustrations
  from articular cartilage, in: G.~A. Holzapfel, R.~W. Ogden (Eds.),
  Biomechanics: Trends in Modeling and Simulation, Springer, 2017, Ch.~1, pp.
  1--51.
\newblock \href {https://doi.org/10.1007/978-3-319-41475-1_1}
  {\path{doi:10.1007/978-3-319-41475-1_1}}.

\bibitem{budday19}
S.~Budday, T.~C. Ovaert, G.~A. Holzapfel, P.~Steinmann, E.~Kuhl, Fifty shades
  of brain: a review on the mechanical testing and modeling of brain tissue,
  Arch. Computat. Methods Eng. (2019) 1--44\href
  {https://doi.org/10.1007/s11831-019-09352-w}
  {\path{doi:10.1007/s11831-019-09352-w}}.

\bibitem{forte17}
A.~E. Forte, S.~M. Gentleman, D.~Dini, On the characterization of the
  heterogeneous mechanical response of human brain tissue, Biomech. Model.
  Mechanobiol. 16~(3) (2017) 907--920.
\newblock \href {https://doi.org/10.1007/s10237-016-0860-8}
  {\path{doi:10.1007/s10237-016-0860-8}}.

\bibitem{hosseinifarid20}
M.~Hosseini-Farid, M.~Ramzanpour, J.~McLean, M.~Ziejewski, G.~Karami, A
  poro-hyper-viscoelastic rate-dependent constitutive modeling for the analysis
  of brain tissues, J. Mech. Behav. Biomed. Mater. 102 (2020) 103475.
\newblock \href {https://doi.org/10.1016/j.jmbbm.2019.103475}
  {\path{doi:10.1016/j.jmbbm.2019.103475}}.

\bibitem{comellas20}
E.~Comellas, S.~Budday, J.-P. Pelteret, G.~A. Holzapfel, P.~Steinmann, Modeling
  the porous and viscous responses of human brain tissue behavior, Comput.
  Methods Appl. Mech. Engrg. 369 (2020) 113128.
\newblock \href {https://doi.org/10.1016/j.cma.2020.113128}
  {\path{doi:10.1016/j.cma.2020.113128}}.

\bibitem{wang19}
R.~Wang, M.~Sarntinoranont, Biphasic analysis of rat brain slices under creep
  indentation shows nonlinear tension-compression behavior, J. Mech. Behav.
  Biomed. Mater. 89 (2019) 1 -- 8.
\newblock \href {https://doi.org/10.1016/j.jmbbm.2018.08.043}
  {\path{doi:10.1016/j.jmbbm.2018.08.043}}.

\bibitem{ehlers14}
W.~Ehlers, A.~Wagner, Multi-component modelling of human brain tissue: a
  contribution to the constitutive and computational description of
  deformation, flow and diffusion processes with application to the invasive
  drug-delivery problem, Comput. Methods Biomech. Biomed. Engin. 18~(8) (2014)
  861 -- 879.
\newblock \href {https://doi.org/10.1080/10255842.2013.853754}
  {\path{doi:10.1080/10255842.2013.853754}}.

\bibitem{forte18}
A.~E. Forte, S.~Galvan, D.~Dini, Models and tissue mimics for brain shift
  simulations, Biomech. Model. Mechanobiol. 17~(1) (2018) 249 -- 261.
\newblock \href {https://doi.org/10.1007/s10237-017-0958-7}
  {\path{doi:10.1007/s10237-017-0958-7}}.

\bibitem{suer2020}
M.~Suer, A.~Abd-Elsayed, Patient with traumatic brain injury, in:
  A.~Abd-Elsayed (Ed.), Guide to the Inpatient Pain Consult, Springer, 2020,
  pp. 429--443.
\newblock \href {https://doi.org/10.1007/978-3-030-40449-9_29}
  {\path{doi:10.1007/978-3-030-40449-9_29}}.

\bibitem{jiang15}
Y.~Jiang, G.~Li, L.-X. Qian, S.~Liang, M.~Destrade, Y.~Cao, Measuring the
  linear and nonlinear elastic properties of brain tissue with shear waves and
  inverse analysis, Biomech. Model. Mechanobiol. 14~(5) (2015) 1119--1128.
\newblock \href {https://doi.org/10.1007/s10237-015-0658-0}
  {\path{doi:10.1007/s10237-015-0658-0}}.

\bibitem{trotta20}
A.~Trotta, J.~M. Clark, A.~McGoldrick, M.~D. Gilchrist, A.~N{\'i}~Annaidh,
  Biofidelic finite element modelling of brain trauma: Importance of the scalp
  in simulating head impact, Int. J. Mech. Sci. 173 (2020) 105448.
\newblock \href {https://doi.org/10.1016/j.ijmecsci.2020.105448}
  {\path{doi:10.1016/j.ijmecsci.2020.105448}}.

\bibitem{tripathi19}
B.~B. Tripathi, D.~Esp{\'i}ndola, G.~F. Pinton, Modeling and simulations of two
  dimensional propagation of shear shock waves in relaxing soft solids, J.
  Comput. Phys. 395 (2019) 205--222.
\newblock \href {https://doi.org/10.1016/j.jcp.2019.06.014}
  {\path{doi:10.1016/j.jcp.2019.06.014}}.

\bibitem{panzer12}
M.~B. Panzer, B.~S. Myers, B.~P. Capehart, C.~R. Bass, Development of a finite
  element model for blast brain injury and the effects of csf cavitation, Ann.
  Biomed. Eng. 40~(7) (2012) 1530--1544.
\newblock \href {https://doi.org/10.1007/s10439-012-0519-2}
  {\path{doi:10.1007/s10439-012-0519-2}}.

\bibitem{deboer94}
R.~De~Boer, Z.~Liu, Plane waves in a semi-infinite fluid saturated porous
  medium, Transp. Porous Med. 16~(2) (1994) 147--173.
\newblock \href {https://doi.org/10.1007/BF00617549}
  {\path{doi:10.1007/BF00617549}}.

\bibitem{deboer95}
R.~De~Boer, Z.~Liu, Propagation of acceleration waves in incompressible
  saturated porous solids, Transp. Porous Med. 21~(2) (1995) 163 -- 173.
\newblock \href {https://doi.org/10.1007/BF00613754}
  {\path{doi:10.1007/BF00613754}}.

\bibitem{deboer96b}
R.~De~Boer, Z.~Liu, Growth and decay of acceleration waves in incompressible
  saturated poroelastic solids, ZAMM Z. angew. Math. Mech. 76~(6) (1996)
  341--347.
\newblock \href {https://doi.org/10.1002/zamm.19960760608}
  {\path{doi:10.1002/zamm.19960760608}}.

\bibitem{diebels96}
S.~Diebels, W.~Ehlers, Dynamic analysis of a fully saturated porous medium
  accounting for geometrical and material non-linearities, Int. J. Numer. Meth.
  Eng. 39~(1) (1996) 81--97.
\newblock \href
  {https://doi.org/10.1002/(SICI)1097-0207(19960115)39:1<81::AID-NME840>3.0.CO;2-B}
  {\path{doi:10.1002/(SICI)1097-0207(19960115)39:1<81::AID-NME840>3.0.CO;2-B}}.

\bibitem{breuer99}
S.~Breuer, Quasi-static and dynamic behavior of saturated porous media with
  incompressible constituents, Transp. Porous Med. 34~(1-3) (1999) 285--303.
\newblock \href {https://doi.org/10.1023/A:1006586130476}
  {\path{doi:10.1023/A:1006586130476}}.

\bibitem{heider12}
Y.~Heider, B.~Markert, W.~Ehlers, Dynamic wave propagation in infinite
  saturated porous media half spaces, Comput. Mech. 49~(3) (2012) 319--336.
\newblock \href {https://doi.org/10.1007/s00466-011-0647-9}
  {\path{doi:10.1007/s00466-011-0647-9}}.

\bibitem{ciarletta18}
M.~Ciarletta, B.~Straughan, V.~Tibullo, Acceleration waves in a nonlinear biot
  theory of porous media, Int. J. Non-Linear Mech. 103 (2018) 23 -- 26.
\newblock \href {https://doi.org/10.1016/j.ijnonlinmec.2018.04.005}
  {\path{doi:10.1016/j.ijnonlinmec.2018.04.005}}.

\bibitem{connolly15}
J.~A.~D. Connolly, Y.~Y. Podladchikov, An analytical solution for solitary
  porosity waves: dynamic permeability and fluidization of nonlinear viscous
  and viscoplastic rock, Geofluids 15~(1-2) (2015) 269--292.
\newblock \href {https://doi.org/10.1111/gfl.12110}
  {\path{doi:10.1111/gfl.12110}}.

\bibitem{bourbie87}
T.~Bourbi{\'e}, O.~Coussy, B.~Zinszner, Acoustics of porous media, {\'E}ditions
  Technip, 1987.

\bibitem{carcione15}
J.~M. Carcione, Wave Fields in Real Media, 3rd Edition, Elsevier Science, 2015.
\newblock \href {https://doi.org/10.1016/C2013-0-18893-9}
  {\path{doi:10.1016/C2013-0-18893-9}}.

\bibitem{grinfeld96}
M.~A. Grinfeld, A.~N. Norris, Acoustoelasticity theory and applications for
  fluid-saturated porous media, J. Acoust. Soc. Am. 100~(3) (1996) 1368 --
  1374.
\newblock \href {https://doi.org/10.1121/1.415983}
  {\path{doi:10.1121/1.415983}}.

\bibitem{tong17}
L.~H. Tong, Y.~S. Liu, D.~X. Geng, S.~K. Lai, Nonlinear wave propagation in
  porous materials based on the biot theory, J. Acoust. Soc. Am. 142~(2) (2017)
  756--770.
\newblock \href {https://doi.org/10.1121/1.4996439}
  {\path{doi:10.1121/1.4996439}}.

\bibitem{deboer05}
R.~De~Boer, Trends in Continuum Mechanics of Porous Media, Springer, 2005.
\newblock \href {https://doi.org/10.1007/1-4020-3144-0}
  {\path{doi:10.1007/1-4020-3144-0}}.

\bibitem{coussy04}
O.~Coussy, Poromechanics, John Wiley \& Sons, Ltd, 2004.
\newblock \href {https://doi.org/10.1002/0470092718}
  {\path{doi:10.1002/0470092718}}.

\bibitem{bowen76}
R.~M. Bowen, Theory of mixtures, in: A.~C. Eringen (Ed.), Continuum Physics,
  Vol. III, Academic Press, 1976, pp. 1--127.

\bibitem{rajagopal95}
K.~R. Rajagopal, L.~Tao, Mechanics of Mixtures, World Scientific, 1995.
\newblock \href {https://doi.org/10.1142/2197} {\path{doi:10.1142/2197}}.

\bibitem{ehlers02}
W.~Ehlers, Foundations of multiphasic and porous materials, in: W.~Ehlers,
  J.~Bluhm (Eds.), Porous Media, Springer, 2002.
\newblock \href {https://doi.org/10.1007/978-3-662-04999-0_1}
  {\path{doi:10.1007/978-3-662-04999-0_1}}.

\bibitem{macminn16}
C.~W. MacMinn, E.~R. Dufresne, J.~S. Wettlaufer, Large deformations of a soft
  porous material, Phys. Rev. Applied 5~(4) (2016) 044020.
\newblock \href {https://doi.org/10.1103/PhysRevApplied.5.044020}
  {\path{doi:10.1103/PhysRevApplied.5.044020}}.

\bibitem{romenski20}
E.~Romenski, G.~Reshetova, I.~Peshkov, M.~Dumbser, Modeling wavefields in
  saturated elastic porous media based on thermodynamically compatible system
  theory for two-phase solid-fluid mixtures, Comput. Fluids (2020) 104587\href
  {https://doi.org/10.1016/j.compfluid.2020.104587}
  {\path{doi:10.1016/j.compfluid.2020.104587}}.

\bibitem{godunov03}
S.~K. Godunov, E.~Romenskii, Elements of Continuum Mechanics and Conservation
  Laws, Springer, 2003.
\newblock \href {https://doi.org/10.1007/978-1-4757-5117-8}
  {\path{doi:10.1007/978-1-4757-5117-8}}.

\bibitem{wilmanski05}
K.~Wilmanski, Tortuosity and objective relative accelerations in the theory of
  porous materials, Proc. R. Soc. A 461~(2057) (2005) 1533--1561.
\newblock \href {https://doi.org/10.1098/rspa.2004.1423}
  {\path{doi:10.1098/rspa.2004.1423}}.

\bibitem{pierce19}
A.~D. Pierce, Acoustics, 3rd Edition, Springer, 2019.
\newblock \href {https://doi.org/10.1007/978-3-030-11214-1}
  {\path{doi:10.1007/978-3-030-11214-1}}.

\bibitem{jordan17}
P.~M. Jordan, F.~Passarella, V.~Tibullo, Poroacoustic waves under a
  mixture-theoretic based reformulation of the jordan–darcy–cattaneo model,
  Wave Motion 71 (2017) 82 -- 92.
\newblock \href {https://doi.org/10.1016/j.wavemoti.2016.07.014}
  {\path{doi:10.1016/j.wavemoti.2016.07.014}}.

\bibitem{markert07}
B.~Markert, A constitutive approach to 3-d nonlinear fluid flow through finite
  deformable porous continua, Transp. Porous Med. 70~(3) (2007) 427.
\newblock \href {https://doi.org/10.1007/s11242-007-9107-6}
  {\path{doi:10.1007/s11242-007-9107-6}}.

\bibitem{franceschini06}
G.~Franceschini, D.~Bigoni, P.~Regitnig, G.~A. Holzapfel, Brain tissue deforms
  similarly to filled elastomers and follows consolidation theory, J. Mech.
  Phys. Solids 54~(12) (2006) 2592--2620.
\newblock \href {https://doi.org/10.1016/j.jmps.2006.05.004}
  {\path{doi:10.1016/j.jmps.2006.05.004}}.

\bibitem{li04}
C.~Li, R.~I. Borja, R.~A. Regueiro, Dynamics of porous media at finite strain,
  Comput. Methods Appl. Mech. Engrg. 193~(36-38) (2004) 3837--3870.
\newblock \href {https://doi.org/10.1016/j.cma.2004.02.014}
  {\path{doi:10.1016/j.cma.2004.02.014}}.

\bibitem{auriault09}
J.-L. Auriault, C.~Boutin, C.~Geindreau, Homogenization of Coupled Phenomena in
  Heterogenous Media, ISTE Ltd, 2009.
\newblock \href {https://doi.org/10.1002/9780470612033}
  {\path{doi:10.1002/9780470612033}}.

\bibitem{sciarra07}
G.~Sciarra, F.~dell'Isola, O.~Coussy, Second gradient poromechanics, Int. J.
  Solids Struct. 44~(20) (2007) 6607--6629.
\newblock \href {https://doi.org/10.1016/j.ijsolstr.2007.03.003}
  {\path{doi:10.1016/j.ijsolstr.2007.03.003}}.

\bibitem{depascalis19}
R.~De~Pascalis, G.~Napoli, G.~Saccomandi, Kink-type solitary waves within the
  quasi-linear viscoelastic model, Wave Motion 86 (2019) 195--202.
\newblock \href {https://doi.org/10.1016/j.wavemoti.2018.12.004}
  {\path{doi:10.1016/j.wavemoti.2018.12.004}}.

\bibitem{johnson82}
D.~L. Johnson, T.~J. Plona, Acoustic slow waves and the consolidation
  transition, J. Acoust. Soc. Am. 72~(2) (1982) 556--565.
\newblock \href {https://doi.org/10.1121/1.388036}
  {\path{doi:10.1121/1.388036}}.

\bibitem{nayfeh00}
A.~H. Nayfeh, Perturbation Methods, WILEY‐VCH Verlag GmbH {\&} Co. KGaA,
  2000.
\newblock \href {https://doi.org/10.1002/9783527617609}
  {\path{doi:10.1002/9783527617609}}.

\bibitem{ruggeri98}
I.~M{\"u}ller, T.~Ruggeri, Rational Extended Thermodynamics, 2nd Edition,
  Springer, 1998.
\newblock \href {https://doi.org/10.1007/978-1-4612-2210-1}
  {\path{doi:10.1007/978-1-4612-2210-1}}.

\bibitem{godlewski96}
E.~Godlewski, P.-A. Raviart, Numerical Approximation of Hyperbolic Systems of
  Conservation Laws, Springer, 1996.
\newblock \href {https://doi.org/10.1007/978-1-4612-0713-9}
  {\path{doi:10.1007/978-1-4612-0713-9}}.

\end{thebibliography}

\appendix

%%%%%%%%%%%%% Appendix %%%%%%%%%%%%%%%

\section{System matrices}\label{app:Mat}

Using the invariance assumption along $y$, $z$, the deformation gradient and distorsion tensors can be simplified, see Eq.~\eqref{F1D}. Moreover, the system \eqref{HypSmall} becomes
\begin{equation}
	\left\lbrace
	\begin{aligned}
		&\partial_t A_{i1} + \partial_x( A_{ik} {v}^\text{s}_{k}) = 0, \\
		&\partial_t n^\text{f} + \partial_x \big( n^\text{f} (w_{1} + v^\text{s}_{1}) \big) = 0, \\
		& \partial_x \big( n^\text{f} w_{1} + v^\text{s}_{1}\big) = 0, \\
		&\rho^\text{f} (\partial_t v^\text{s}_i + v^\text{f}_1 \partial_x v^\text{s}_{i}) + a \rho^\text{f} (\partial_t w_i + v^\text{f}_1 \partial_x w_i) + \delta_{i1} n^\text{f}\partial_x p = -\tfrac{(n^\text{f})^2}{k^\text{f}} w_i, \\
		&\rho\, (\partial_t v^\text{s}_i + v_1 \partial_x v^\text{s}_i) + \rho^\text{f} (\partial_t w_i + v^\text{f}_1 \partial_x w_i) - \partial_x \sigma^\text{e}_{i1} + \delta_{i1} \partial_x p = 0,
	\end{aligned}
	\right.
	\label{Hyp1D}
\end{equation}
for indices $i$ ranging from one to three, where Einstein's notation for repeated indices was used. Here, we have used the notation $\rho\bm{v} = \rho^\text{s}\bm{v}^\text{s} + \rho^\text{f}\bm{v}^\text{f}$ for the mixture momentum, where $\rho = \rho^\text{s} + \rho^\text{f}$ denotes the mixture density.

Let us rewrite this system in quasi-linear form \eqref{Hyp1D}. To do so, we expand the spatial derivatives by using the product rule. By virtue of the chain rule, the spatial derivatives of the stress components $\bm{\sigma}^\text{e}$ become $\partial_x {\sigma}^\text{e}_{i1} = -Q_{ij} \partial_x A_{j1}$ with the coefficients $Q_{ij} = -\partial {\sigma}^\text{e}_{i1}/\partial A_{j1}$.
Considering the unidimensional deformation defined in Eq.~\eqref{F1D}, the constitutive law \eqref{NeoHookean} for the solid skeleton gives
\begin{equation}
	\begin{aligned}
	\bm{\sigma}^\text{e} &= \frac{\mu}{A_{11}} \begin{bmatrix}
		1 - A_{11}^2(1 + \gamma \ln A_{11}) & - A_{21}  & - A_{31} \\
		- A_{21} &  A_{21}^2 - \gamma A_{11}^2 \ln A_{11} &  A_{21} A_{31} \\
		- A_{31} &  A_{21} A_{31} &  A_{31}^2 - \gamma A_{11}^2 \ln A_{11}
	\end{bmatrix} \\
	&\quad + \mu \beta \frac{1 + A_{21}^2 + A_{31}^2 - A_{11}^2}{A_{11}^3} \begin{bmatrix}
		1 & -A_{21} & -A_{31} \\
		-A_{21} & A_{11}^2+A_{21}^2 & A_{21} A_{31} \\
		-A_{31}  & A_{21} A_{31}  & A_{11}^2+A_{31}^2
	\end{bmatrix}
	\end{aligned}
	\label{Terzaghi1D}
\end{equation}
with $\gamma = \lambda/\mu$.
The coefficients $Q_{ij} = -\partial {\sigma}^\text{e}_{i1}/\partial A_{j1}$ are therefore given by
\begin{equation*}
	\begin{aligned}
	&\qquad\qquad\qquad \left[Q_{ij}\right] = \frac{\mu}{A_{11}} \begin{bmatrix}
		A_{11}^{-1} + (1 + \gamma + \gamma\ln A_{11}) A_{11} & 0 & 0 \\
		-A_{21}/A_{11} & 1 & 0 \\
		-A_{31}/A_{11} & 0 & 1
	\end{bmatrix} \\
	&+ \frac{\mu \beta}{A_{11}^4}
	{\setlength{\arraycolsep}{1pt}
	\begin{bmatrix}
		3-A_{11}^2+3 A_{21}^2+3 A_{31}^2 & -2 A_{11} A_{21} & -2 A_{11} A_{31} \\
		(A_{11}^2-3 A_{21}^2-3 A_{31}^2-3) A_{21} & (1-A_{11}^2+3 A_{21}^2+A_{31}^2) A_{11} & 2 A_{11}  A_{21} A_{31} \\
		(A_{11}^2-3 A_{21}^2-3 A_{31}^2-3) A_{31} & 2 A_{11} A_{21} A_{31} & (1-A_{11}^2+A_{21}^2+3 A_{31}^2) A_{11}
	\end{bmatrix}}
	\end{aligned}
\end{equation*}
which may be viewed as deformation-dependent elastic moduli. In fact, in the limit of Hookean linear elasticity (or equivalently, in a static undeformed state), the only non-zero coefficients $Q_{ij}$ are $Q_{11} \simeq \lambda + 2 \mu$ and $Q_{22} = Q_{33} \simeq \mu$ where $\lambda$, $\mu$ are the Lam{\'e} parameters.
Finally, we end up with the quasi-linear first-order system \eqref{Hyp}, with the $11 \times 11$ matrices ${\bf M}^\nu$ and the vector $\bf R$ specified below.
\begin{equation*}
	{\bf M}^t = 
	{\setlength{\arraycolsep}{2pt}
		\setcounter{MaxMatrixCols}{11}
		\begin{bmatrix}
			1 & 0 & 0 & 0 & 0 & 0 & 0 & 0 & 0 & 0 & 0\\
			0 & 1 & 0 & 0 & 0 & 0 & 0 & 0 & 0 & 0 & 0\\
			0 & 0 & 1 & 0 & 0 & 0 & 0 & 0 & 0 & 0 & 0\\
			0 & 0 & 0 & 1 & 0 & 0 & 0 & 0 & 0 & 0 & 0\\
			0 & 0 & 0 & 0 & \rho^\text{f} & 0 & 0 & a \rho^\text{f} & 0 & 0 & 0\\
			0 & 0 & 0 & 0 & 0 & \rho^\text{f} & 0 & 0 & a \rho^\text{f} & 0 & 0\\
			0 & 0 & 0 & 0 & 0 & 0 & \rho^\text{f} & 0 & 0 & a \rho^\text{f} & 0\\
			0 & 0 & 0 & 0 & \rho & 0 & 0 & \rho^\text{f} & 0 & 0 & 0\\
			0 & 0 & 0 & 0 & 0 & \rho & 0 & 0 & \rho^\text{f} & 0 & 0\\
			0 & 0 & 0 & 0 & 0 & 0 & \rho & 0 & 0 & \rho^\text{f} & 0\\
			0 & 0 & 0 & 0 & 0 & 0 & 0 & 0 & 0 & 0 & 0
	\end{bmatrix}}
	,
	\label{Mt}
\end{equation*}
\begin{equation*}
	{\bf M}^x = 
	{\setlength{\arraycolsep}{2pt}
		\setcounter{MaxMatrixCols}{11}
		\begin{bmatrix}
			v^\text{s}_1 & 0 & 0 & 0 & A_{11} & 0 & 0 & 0 & 0 & 0 & 0\\
			0 & v^\text{s}_1 & 0 & 0 & A_{21} & 1 & 0 & 0 & 0 & 0 & 0\\
			0 & 0 & v^\text{s}_1 & 0 & A_{31} & 0 & 1 & 0 & 0 & 0 & 0\\
			0 & 0 & 0 & v^\text{f}_1 & n^\text{f} & 0 & 0 & n^\text{f} & 0 & 0 & 0\\
			0 & 0 & 0 & 0 & \rho^\text{f} v^\text{f}_1 & 0 & 0 & a \rho^\text{f} v^\text{f}_1 & 0 & 0 & n^\text{f}\\
			0 & 0 & 0 & 0 & 0 & \rho^\text{f} v^\text{f}_1 & 0 & 0 & a \rho^\text{f} v^\text{f}_1 & 0 & 0\\
			0 & 0 & 0 & 0 & 0 & 0 & \rho^\text{f} v^\text{f}_1 & 0 & 0 & a \rho^\text{f} v^\text{f}_1 & 0\\
			Q_{11} & Q_{12} & Q_{13} & 0 & \rho v_1 & 0 & 0 & \rho^\text{f} v^\text{f}_1 & 0 & 0 & 1\\
			Q_{21} & Q_{22} & Q_{23} & 0 & 0 & \rho v_1 & 0 & 0 & \rho^\text{f} v^\text{f}_1 & 0 & 0\\
			Q_{31} & Q_{32} & Q_{33} & 0 & 0 & 0 & \rho v_1 & 0 & 0 & \rho^\text{f} v^\text{f}_1 & 0\\
			0 & 0 & 0 & w_1 & 1 & 0 & 0 & n^\text{f} & 0 & 0 & 0
	\end{bmatrix}}, \qquad
	%\end{equation*}
	%\begin{equation*}
	{\bf R} = -\frac{(n^\text{f})^2}{k^\text{f}}
	\begin{bmatrix}
		0\\
		0\\
		0\\
		0\\
		w_1\\
		w_2\\
		w_3\\
		0\\
		0\\
		0\\
		0
	\end{bmatrix} .
\end{equation*}

\end{document}